\documentclass[prl,aps,showpacs,groupedaddress,superscriptaddress,twocolumn,toc=flat]{revtex4-2}

\usepackage[T1]{fontenc}

\usepackage{physics}
\usepackage{amsmath,amssymb}
\usepackage{graphicx}
\usepackage[dvipsnames]{xcolor}
\usepackage[caption=false]{subfig}
\usepackage[colorlinks=true,linkcolor=blue,urlcolor=blue,citecolor=blue]{hyperref}
\usepackage{nicefrac}

\begin{document}
\title{Snapshot-based detection of $\nicefrac{1}{2}$-Laughlin states: coupled chains and central charge}

\author{F. A. Palm}
\affiliation{Department of Physics and Arnold Sommerfeld Center for Theoretical Physics (ASC), Ludwig-Maximilians-Universit\"at M\"unchen, Theresienstr. 37, D-80333 M\"unchen, Germany}
\affiliation{Munich Center for Quantum Science and Technology (MCQST), Schellingstr. 4, D-80799 M\"unchen, Germany}

\author{S. Mardazad}
\affiliation{Department of Physics and Arnold Sommerfeld Center for Theoretical Physics (ASC), Ludwig-Maximilians-Universit\"at M\"unchen, Theresienstr. 37, D-80333 M\"unchen, Germany}
\affiliation{Munich Center for Quantum Science and Technology (MCQST), Schellingstr. 4, D-80799 M\"unchen, Germany}

\author{A. Bohrdt}
\affiliation{Department of Physics, Harvard University, Cambridge, MA 02138, USA}
\affiliation{ITAMP, Harvard-Smithsonian Center for Astrophysics, Cambridge, MA 02138, USA}

\author{U. Schollw\"ock}
\affiliation{Department of Physics and Arnold Sommerfeld Center for Theoretical Physics (ASC), Ludwig-Maximilians-Universit\"at M\"unchen, Theresienstr. 37, D-80333 M\"unchen, Germany}
\affiliation{Munich Center for Quantum Science and Technology (MCQST), Schellingstr. 4, D-80799 M\"unchen, Germany}

\author{F. Grusdt}
\affiliation{Department of Physics and Arnold Sommerfeld Center for Theoretical Physics (ASC), Ludwig-Maximilians-Universit\"at M\"unchen, Theresienstr. 37, D-80333 M\"unchen, Germany}
\affiliation{Munich Center for Quantum Science and Technology (MCQST), Schellingstr. 4, D-80799 M\"unchen, Germany}

\date{\today}

\begin{abstract}
	Experimental realizations of topologically ordered states of matter, such as fractional quantum Hall states, with cold atoms are now within reach.
	In particular, optical lattices provide a promising platform for the realization and characterization of such states, where novel detection schemes enable an unprecedented microscopic understanding.
	Here we show that the central charge can be directly measured in current cold atom experiments using the number entropy as a proxy for the entanglement entropy.
	We perform density-matrix renormalization-group simulations of Hubbard-interacting bosons on coupled chains subject to a magnetic field with $\alpha=\nicefrac{1}{4}$ flux quanta per plaquette.
	Tuning the inter-chain hopping, we find a transition from a trivial quasi-one dimensional phase to the topologically ordered Laughlin state at magnetic filling factor $\nu=\nicefrac{1}{2}$ for systems of three or more chains.
	We resolve the transition using the central charge, on-site correlations, momentum distributions and the many-body Chern number.
	Additionally, we propose a scheme to experimentally estimate the central charge from Fock basis snapshots.
	The model studied here is experimentally realizable with existing cold atom techniques and the proposed observables pave the way for the detection and classification of a larger class of interacting topological states of matter.
\end{abstract}

\maketitle

\paragraph{Introduction.---}
The interplay of topological band structures and interactions has been a fruitful source of exotic quantum phases.
Most prominently, the fractional quantum Hall (FQH) effect~\cite{Stormer1983} can be understood in the framework of topologically ordered phases of matter.
For example, Laughlin's successful trial wave functions~\cite{Laughlin1983} are known to have excitations exhibiting Abelian anyonic braiding~\cite{Halperin1984,Arovas1984}.
Up to now, the FQH effect is best studied in solid-state experiments, but proposals and first implementations of alternative realizations exist~\cite{Soerensen2005,Palmer2006,Hafezi2007,Cho2008,Hayward2012,Hormozi2012,Cooper2013,Yao2013,Hafezi2013,Kapit2014,Sterdyniak2015a,Anderson2016,Lacki2016,Taddia2017}.

Early cold atom experiments used rotating traps to mimic the effect of the external magnetic field needed to reach the FQH regime~\cite{AboShaeer2001,Schweikhard2004,Fletcher2021}.
In this setup, first signatures of the bosonic Laughlin state at $\nu=\frac{1}{2}$ have been observed~\cite{Gemelke2010}.
A high degree of control and flexibility as well as site-resolved imaging techniques make cold atoms in optical lattices a promising platform to further study the correlated nature of FQH states.
Extensive numerical studies have found evidence for various FQH states in experimentally realistic models like the Hofstadter-Bose-Hubbard (HBH) model~\cite{Soerensen2005,Hafezi2007,Palmer2008,Moeller2009,Moeller2015,Bauer2016,Huegel2017,Motruk2017,He2017,Gerster2017,Andrews2018,Dong2018,Andrews2021,Palm2021,Andrews2021a,Boesl2022,Wang2022}.
In recent years, experimental progress led to the implementation of non-interacting~\cite{Aidelsburger2013,Miyake2013,Stuhl2015,Mancini2015} and interacting~\cite{Tai2017} Hofstadter models using ultracold atoms in optical lattices and has paved the way towards cold atom realizations of FQH states in the very near future.
This includes models with anisotropic hopping, where approaches starting from decoupled, one-dimensional chains provide a promising route towards the successful adiabatic preparation of topologically ordered states~\cite{He2017}.
However, viable experimental schemes for elucidating the topological nature of the states remain scarce.

Here, we study Hubbard-interacting bosons subject to a magnetic field at filling factor $\nu=\nicefrac{1}{2}$ on chains with tunable inter-chain hopping.
We perform density-matrix renormalization-group (DMRG) simulations to calculate the ground state of the HBH model for varying spatial anisotropy.
While finite-size and lattice effects affect microscopic properties of the wave function, we focus on universal properties which are robust to such effects.
In particular, we use the central charge to identify the ground state close to the isotropic limit in systems with three or more chains as a lattice analog of the $\nicefrac{1}{2}$-Laughlin state.
Furthermore, we propose a scheme to extract the central charge from snapshots in current cold atom experiments with quantum gas microscopes.
We also discuss signatures of the topological phase in experimentally more established observables like the momentum distribution along the chains.
Finally, we clarify the topological nature of the ground state by calculating the many-body Chern number as function of the inter-chain hopping strength.

The approach pursued here is inspired by analytical coupled wire constructions in the continuum~\cite{Kane2002,Teo2014,Fuji2019}, describing how inter-chain coupling can lead to the formation of topologically ordered states.
A prime example is the emergence of Laughlin and hierarchy states at proper filling factors.
Furthermore, coupled wires allow for an intuitive understanding of the structure of the edge theory as well as the quasi-particles of such states.
The generalization of this approach to discrete chains provides a promising way to construct exotic quantum phases in an experimentally accessible setup~\cite{He2017}, and motivates our probes of topological order.

\paragraph{Model.---} We study a lattice version of the bosonic FQH problem in the HBH model on a square lattice of size $L_x \times L_y$.
In particular, we allow for anisotropic hopping, such that the Hamiltonian in Landau gauge reads
\begin{widetext}
\begin{equation}
\hat{\mathcal{H}} = - t_x \sum_{x=1}^{L_x-1} \sum_{y=1}^{L_y} \left(\hat{a}^{\dagger}_{x+1, y}\hat{a}^{\vphantom{\dagger}}_{x,y} + \mathrm{H.c.}\right) - t_y \sum_{x=1}^{L_x} \sum_{y=1}^{L_y-1} \left(\mathrm{e}^{2\pi i \alpha x}\hat{a}^{\dagger}_{x,y+1}\hat{a}^{\vphantom{\dagger}}_{x,y} + \mathrm{H.c.}\right) +\frac{U}{2}\sum_{x,y} \hat{n}_{x,y}\left(\hat{n}_{x,y}-1\right).
\label{Eq:HBH-Hamiltonian}
\end{equation}
\end{widetext}
Here, $\hat{a}^{(\dagger)}_{x,y}$ are bosonic annihilation (creation) operators and $\hat{n}_{x,y} = \hat{a}_{x,y}^{\dagger}\hat{a}^{\vphantom{\dagger}}_{x,y}$ are the boson number operators.
The first two terms describe (potentially anisotropic) hopping between neighboring sites, while the last term describes repulsive ($U/t_x > 0$) on-site interactions.
We choose open boundary conditions in both directions and fix the Hubbard interaction strength to $U/t_x = 5$, which is large compared to the band width of the lowest band and also the band gap of the single-particle model.

Furthermore, we restrict ourselves to a magnetic flux per plaquette of $\alpha = N_{\phi} / \left[\left(L_x-1\right)\left(L_y-1\right)\right] = \nicefrac{1}{4}$, so that in the isotropic case, $t_y/t_x=1$, continuum limit analogies of earlier studies~\cite{Soerensen2005} apply.
Thus, at the magnetic filling factor $\nu = \nicefrac{N}{N_{\phi}} = \nicefrac{1}{2}$ studied here, we expect the ground state in the isotropic limit to be closely related to the topologically ordered $\nicefrac{1}{2}$-Laughlin state~\cite{Laughlin1983}.
We will see that this behavior is to some extent robust to tuning the inter-chain hopping strength.

We perform DMRG simulations~\cite{White1992,Schollwoeck2005} using the \textsc{SyTen} toolkit~\cite{HubigSyTen} to study systems of varying size using the single-site variant \cite{Hubig2015} and truncating the local Hilbert space to at most $N_{\rm max}=3$ bosons per site, justified by the large value of $U/t_x$.
Compared to a hard-core constraint, our truncation avoids an artificial enhancement of Laughlin physics.

\paragraph{Central charge.---}
\begin{figure}
	\centering
	\includegraphics[width=\linewidth]{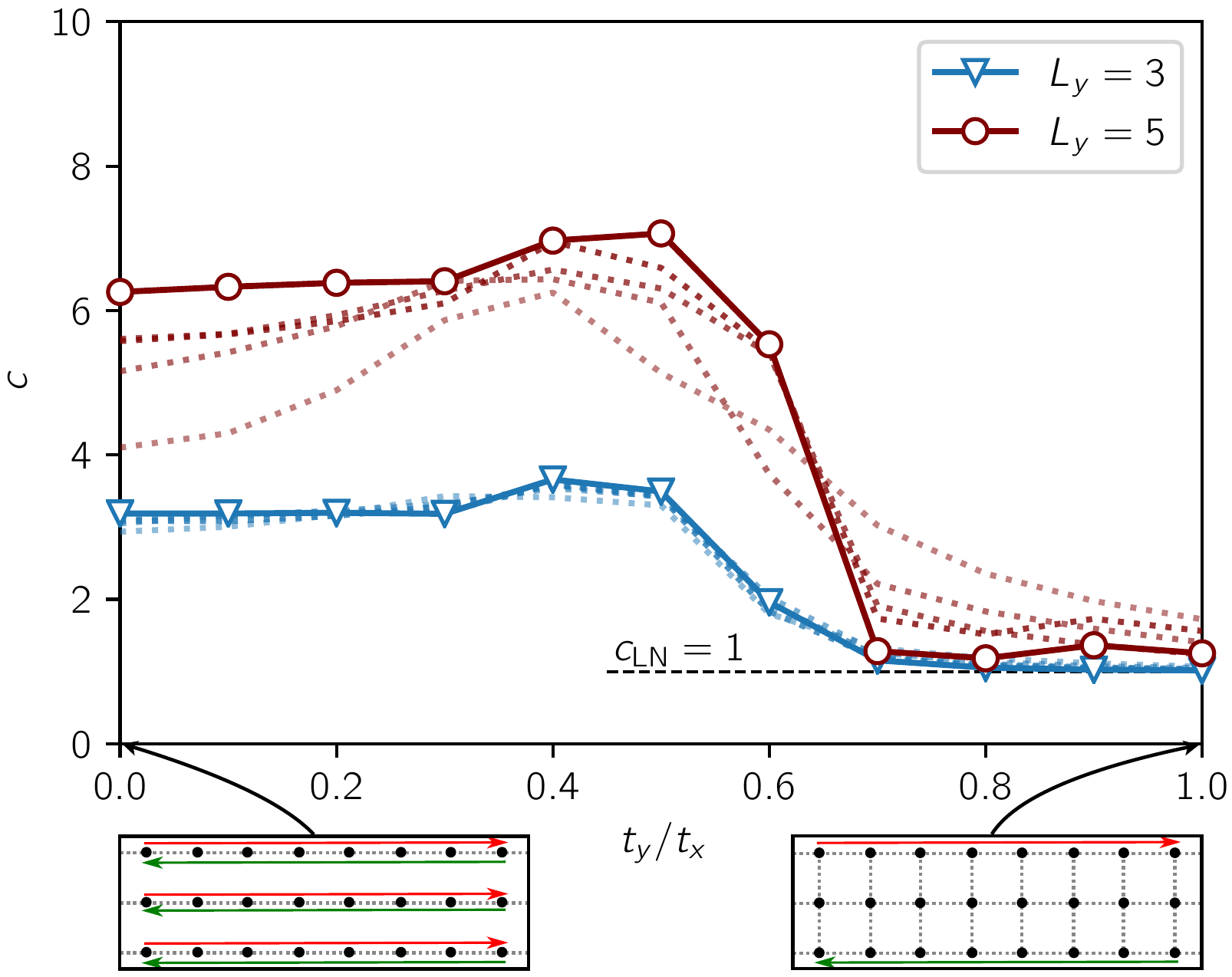}
	\caption{Central charge as obtained from the bipartite entanglement entropy $S(x)$ for different system sizes after extrapolation to $L_x\to\infty$.
		Faded dotted lines indicate values for finite $L_x$ with longer systems being less faded.
		We find a clear change of behavior around $t_y/t_x \approx 0.6$.
		Above this critical value, the central charge is in agreement with the prediction $c_{\rm LN}=1$ for the Laughlin state.
		The sketches below the main panel illustrate the origin of the different behavior in the decoupled ($t_y/t_x=0$) and the isotropic ($t_y/t_x=1$) limit by only showing gapless chiral modes.}
	\label{Fig:CentralCharge}
\end{figure}
Counting the number of chiral gapless modes at the one-dimensional edge, the central charge is an important quantity in the classification of topologically ordered systems.
In particular, it provides a prime quantity to identify FQH states with chiral edge modes.

For the Laughlin state at filling factor $\nu = \nicefrac{1}{2}$, the central charge is predicted to be $c_{\rm LN}=1$.
In our studies, the size of the system along both directions is much larger than the magnetic length so that we expect this prediction to hold true.
Therefore, we expect the central charge to approach unity close to the isotropic limit, $t_y/t_x \approx 1$.
In contrast, in the weakly coupled regime, $t_y/t_x \approx 0$, the system can be considered a collection of $L_y$ independent one-dimensional Luttinger liquids, each of which contributes a value of $c_{\rm LL}=1$ to the total central charge, thus adding up to $c = L_y$.

In order to determine the central charge, we make use of a prediction from conformal field theory (CFT) relating the central charge $c$ to the bipartite entanglement entropy $S(x)$, namely
\begin{equation}
S(x) = \frac{c}{6} \log\left(\frac{2L_x}{\pi} \sin\left(\frac{\pi x}{L_x}\right)\right) + g,
\label{Eq:EntanglementEntropy}
\end{equation}
where $g$ is some non-universal constant and $L_x$ is the length of the system~\cite{Calabrese2004}.

Numerically, the entanglement entropy can be obtained easily from matrix product states (MPS).
By appropriately choosing the MPS chain, the bipartite entanglement entropy between the two parts of the system is entirely carried by a single MPS bond.
Therefore, upon cutting this bond we obtain a bipartition of the underlying lattice along the $x$-direction.
In our finite-size calculations, we account for oscillations in the entanglement entropy, in particular in small systems, by normalizing the entropy to the ambient densities~\cite{supp}.
Furthermore, we extrapolate the entanglement entropy to infinite bond dimensions before we perform a fit using Eq.~\eqref{Eq:EntanglementEntropy} to extract the central charge.
For details of our procedure and additional data points see Ref.~\cite{supp}.

For $L_y\geq 3$-leg systems, we find the value of the central charge to change drastically around $t_y/t_x \approx 0.6$, see Fig.~\ref{Fig:CentralCharge}.
In particular, at large $t_y/t_x$ we find that the numerical central charge almost perfectly matches the theoretical prediction of $c_{\rm LN}=1$.
In the weakly coupled regime convergence of the DMRG calculations is difficult to achieve and the numerical values for the central charge do not coincide with the predicted values to the same degree of accuracy.
Nevertheless, a clear change of behavior is visible in all the systems with $L_y=3,\hdots,6$ chains studied in this work~\cite{supp}.
Given the robustness of the transition with respect to the number of chains, we believe this feature to carry over to the thermodynamic limit and consider it striking evidence for the emergence of a Laughlin phase around $t_y/t_x \approx 1$.

Remarkably, our DMRG simulations rule out a Laughlin-like state in $L_y=2$-leg systems, where we observe $c=2$ throughout~\cite{supp}.
However, additional nearest-neighbor repulsion has been argued to reintroduce Laughlin-like states~\cite{supp}.

\paragraph{Measuring the central charge in experiments.---}
To our knowledge, the central charge has so far eluded direct experimental measurements.
We propose a protocol to measure the central charge in state-of-the-art quantum simulation platforms such as quantum gas microscopes.
The typical outcome of these experiments are projective measurements in the Fock basis resulting in site resolved snapshots of the local particle number.
Efficient methods to generate accurate snapshots from MPS have been developed~\cite{Ferris2012} and proved useful for sampling realistic experimental outcomes in models similar to ours~\cite{Buser2022}.

In order to extract the entanglement entropy $S(x)$ experimentally, we propose to use the particle number entropy $S_n(x)$ as a meaningful proxy in certain regimes.
A similar approach has proven useful in the context of many-body localization~\cite{Lukin2019}, and also theoretical attempts to study the entanglement entropy using particle number fluctuations have been undertaken earlier~\cite{Petrescu2014}.
Now, we exemplify the use of snapshots and their number entropy to determine the central charge of the topologically non-trivial $\nicefrac{1}{2}$-Laughlin state.

The main advantage of the number entropy is that it can be directly extracted from a given set of snapshots.
To this end, each snapshot is split into two subsystems $A$ and $\bar{A}$ and the probability $p_{N_A}$ to observe $N_A$ particles in subsystem $A$ is determined.
Then, the particle number entropy \mbox{$S_n = \sum_{N_A} p_{N_A} \log(p_{N_A})$} can be calculated.
Repeating this scheme for different partitions of the system similar to the case of the entanglement entropy $S$ above, one obtains the number entropy $S_n(x)$ as function of the cut position.

\begin{figure}
	\centering
	\includegraphics[width=\linewidth]{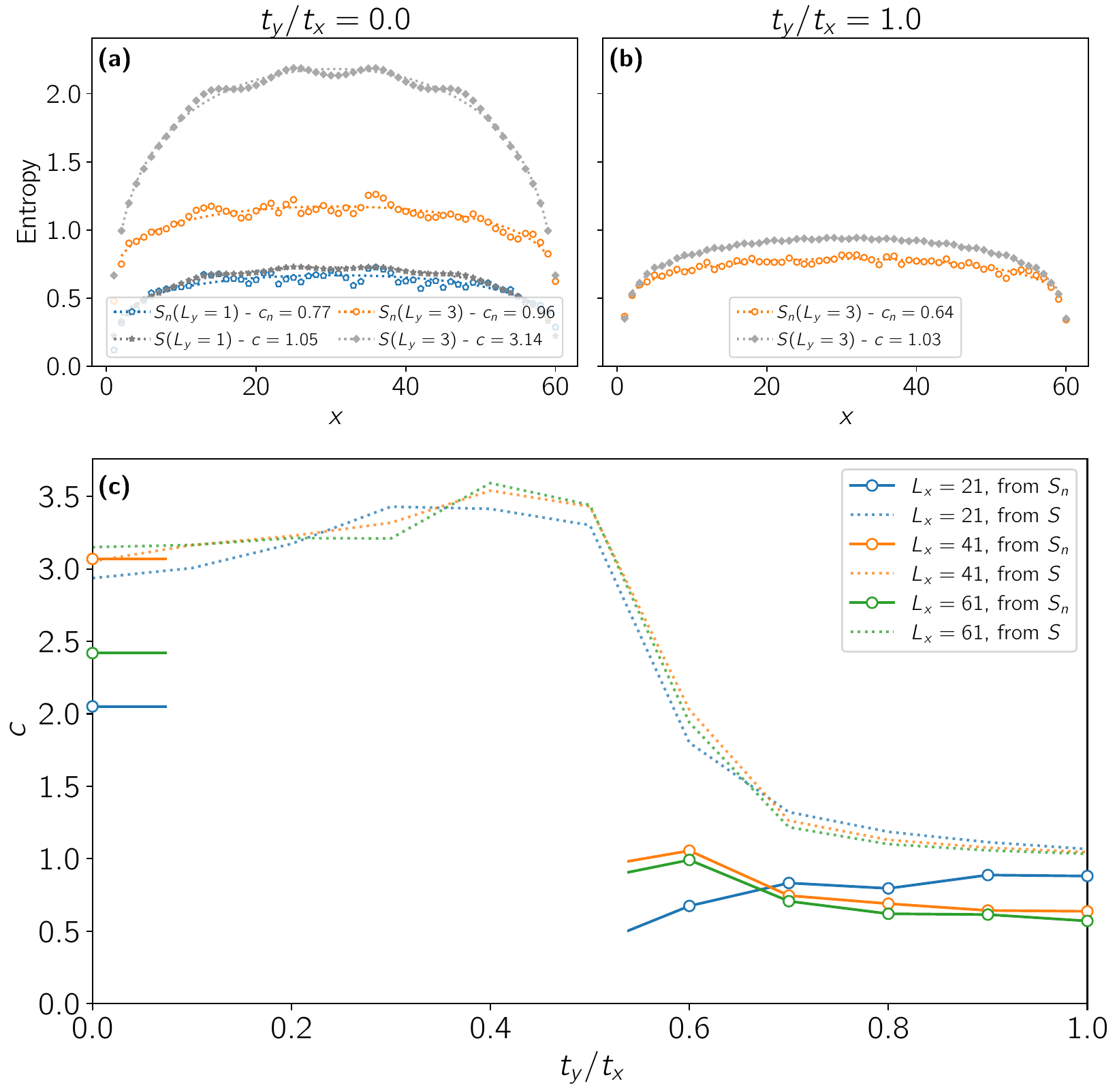}
	\caption{(a) Number entropy $S_n(x)$ from 6000 (2000) snapshots of a single chain (three decoupled chains) of length $L_x=61$.
		While the number entropy provides an accurate proxy for the entanglement entropy $S(L_y=1)$ of a single chain, it is not additive in the number of chains.
		(b) In the isotropic limit, $t_y/t_x = 1$, the proxy $S_n(x)$ from 2000 snapshots of the $3$-leg system is relatively accurate.
		(c) Central charges for $3$-leg systems extracted from the number entropy $S_n(x)$ (solid lines) compared to the prediction from the entanglement entropy $S(x)$ (faded, dotted lines).
	}
	\label{Fig:NumberEntropy}
\end{figure}

In the isotropic limit, $t_y/t_x=1$, the number entropy provides a good estimate of the entanglement entropy, see Fig.~\ref{Fig:NumberEntropy}(b).
This behavior carries over to the entire regime in which we have identified the $\nicefrac{1}{2}$-Laughlin state using the entanglement entropy.
In Fig.~\ref{Fig:NumberEntropy}(c) we find that close to the isotropic limit the prediction of the central charge based on snapshots agrees reasonably well with the prediction from the entanglement entropy.
Thus we conclude that this method can indeed be used to estimate the central charge in the Laughlin phase.
In the decoupled limit, $t_y/t_x = 0$, the number entropy extracted from the full system is not additive in the number of legs, see Fig.~\ref{Fig:NumberEntropy}(a).
In contrast, extracting the number entropy from each leg separately, we find that the central charge $c_n$ in each leg is in agreement with the value from the entanglement entropy, so that multiplying $c_n$ by the number of legs provides the correct overall central charge for the whole system.
In the intermediate regime, we attribute discrepancies between the central charge $c$ and the estimate $c_n$ to the non-applicability of the CFT prediction and to the non-additivity of the number entropy.

We emphasize that the proposed measurement of the central charge is solely based on snapshots in the Fock basis, which are routinely generated in experiments with quantum gas microscopes~\cite{Bakr2009,Sherson2010,Preiss2015,Bergschneider2018}.
The number entropy and the estimated central charge discussed here can be extracted from these snapshots without further experimental efforts and are accessible to existing experiments.

\paragraph{Additional experimental observables.---}
Now, we discuss further observables accessible to quantum gas microscopes.
A well known signature of the $\nicefrac{1}{2}$-Laughlin state, reflecting flux-attachment underlying the formation of composite fermions~\cite{Jain1989}, is a strong suppression of on-site correlations,
\begin{equation}
	g^{(2)}(0) = \frac{1}{2 L_x L_y} \sum_{x,y} \left\langle \hat{n}_{x,y} \left(\hat{n}_{x,y}-1\right) \right\rangle.
\end{equation}
By allowing up to three bosons per site in our numerics, we do not artificially stabilize the Laughlin state by imposing a hard-core constraint for the bosons, formally $U/t =\infty$.
In contrast, given the small particle number densities, our simulations can be expected to properly describe the experimental situation of a finite two-body interaction.

\begin{figure}
	\centering
	\includegraphics[width=\linewidth]{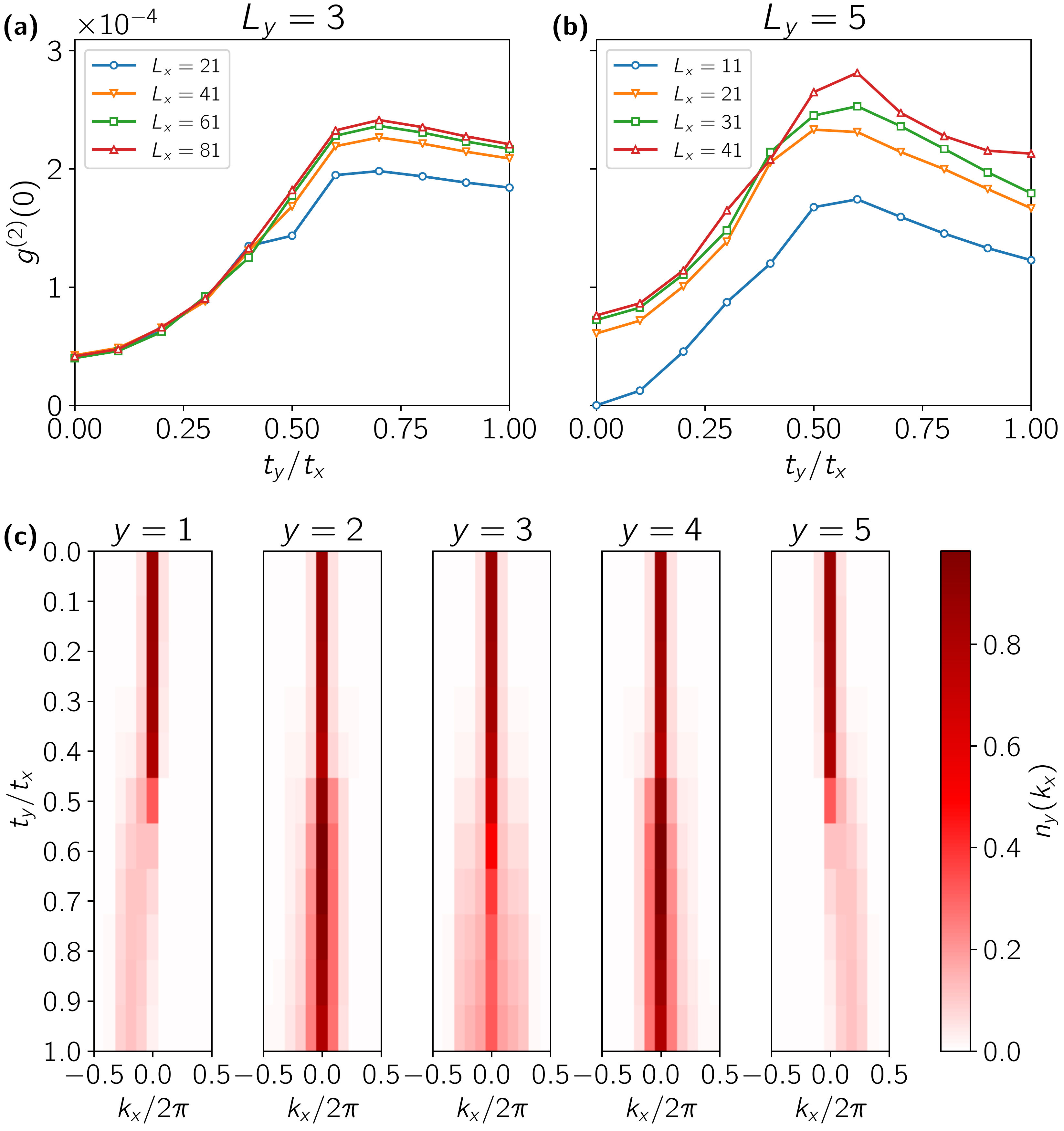}
	\caption{On-site correlations $g^{(2)}(0)$ for (a) $L_y=3$ and (b) $L_y=5$ chains.
		The transition region at intermediate $t_y/t_x$ is clearly visible by a maximum of $g^{(2)}(0)$.
		(c) Momentum distribution $n_y(k_x)$ for $L_y=5, L_x=11, U/t_x=5.0$.
		We observe the emergence of a chiral mode in the outermost chains ($y=1, 5$) around $t_y/t_x \approx 0.6$.
		In these chains, the momentum distribution is peaked around a finite momentum $k_x\neq 0$.
		In the remaining bulk chains, the momentum distribution is peaked around $k_x=0$ at all $t_y/t_x$.}
	\label{Fig:ExperimentalObservables}
\end{figure}

In Fig.~\ref{Fig:ExperimentalObservables}(a,b) we show $g^{(2)}(0)$ as a function of $t_y/t_x$.
In particular, we find that in the weakly coupled limit $g^{(2)}(0)$ is very small, while it increases with increasing inter-chain hopping.
$g^{(2)}(0)$ takes a global maximum around $t_y/t_x \approx 0.6$ before it decreases again in the strongly coupled regime.

In the isotropic limit, this drop is a key feature of the $\nicefrac{1}{2}$-Laughlin state, indicating the screened interactions of composite fermions~\cite{Jain1989}.
In contrast, in the decoupled limit this is the result of a different Jordan-Wigner-type fractionalization of the bosons~\cite{Lieb1963,Lieb1963a,Cheon1999}.
In the intermediate regime, the bosons are not able to fermionize and therefore the residual two-particle correlations are significantly larger compared to both of the limits.

As another experimental observable, we consider the momentum distribution along $x$ in a given wire,
\begin{equation}
	n_{y}(k_x) = \frac{1}{L_x} \sum_{x,x^{\prime}} \mathrm{e}^{-i k_x \left(x-x^{\prime}\right)} \left\langle \hat{a}_{x,y}^{\dagger} \hat{a}_{x^{\prime}, y}^{\vphantom\dagger}\right\rangle,
\end{equation}
where $k_x = \frac{2\pi m}{L_x}$, $m = 0, \hdots, L_x-1$.
The numerical results in Fig.~\ref{Fig:ExperimentalObservables}(c) indicate the occupation of a chiral mode with finite momentum $k_x \neq 0$ at the boundary for $t_y/t_x \gtrsim 0.6$, while in the bulk the distribution remains peaked around $k_x=0$ even around $t_y/t_x \approx 1$.
The momentum distribution in a specific wire is experimentally accessible in various cold atom experiments with current techniques, for example by time-of-flight measurements.

We interpret the sudden change around $t_y/t_x \approx 0.6$ as further evidence for the $\nicefrac{1}{2}$-Laughlin phase close to the isotropic limit indicated by the characteristic chiral edge mode.
This mode also manifests itself in the presence of a chiral edge current for which we also find numerical evidence in our simulations, see Ref.~\cite{supp}.

\paragraph{Topological classification: Many-body Chern number.---}
In order to provide an unambiguous topological classification of the ground state, we determine the many-body Chern number~\cite{Niu1985,Kohmoto1985} as function of the anisotropic hopping strength.
In particular, we use the method proposed by \mbox{Dehghani~\emph{et~al.}~\cite{Dehghani2021}} to extract the many-body Chern number from a single ground state wave function.
To this end, we perform DMRG calculations on cylinders of coupled chains with periodic boundary conditions in $y$-direction~\cite{supp}.

\begin{figure}
	\centering
	\includegraphics[width=\linewidth]{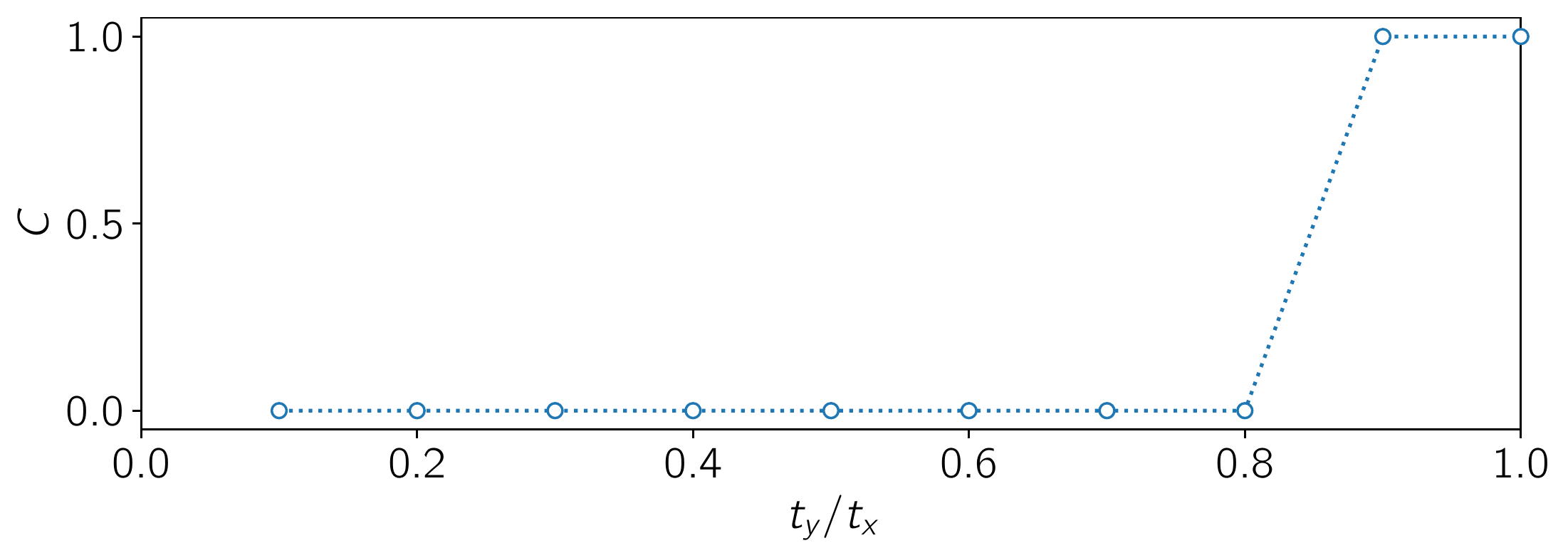}
	\caption{Many-body Chern number as function of $t_y/t_x$ for a cylinder of size $L_x \times L_y = 25\times3$.
		We find a transition towards a topologically non-trivial phase close to the isotropic limit.
	}
	\label{Fig:ChernNumber}
\end{figure}
In Fig.~\ref{Fig:ChernNumber}, we find that for weakly coupled chains the many-body Chern number vanishes, confirming our understanding of the topologically trivial phase in this regime.
On the other hand, around $t_y/t_x \approx 1$ we find the many-body Chern number to be $C=1$, hence resulting in the non-trivial Hall response $\sigma_{\rm H} = \frac{1}{2} \frac{e^2}{h}$ expected for the $\nicefrac{1}{2}$-Laughlin state.
This is in agreement with the other observables discussed here and gives direct evidence for the topological nature of the ground state close to the isotropic limit.

\paragraph{Conclusions.---}
The bosonic $\nicefrac{1}{2}$-Laughlin state can be realized in three or more coupled chains subject to a magnetic field close the isotropic limit.
The transition from a topologically trivial phase to this topologically ordered phase as the inter-chain hopping strength is tuned can be seen from various observables.
Most prominently, we have found the central charge to provide clear evidence in the strong-coupling phase by dropping to the expected value $c_{\rm LN}=1$ for the Laughlin state.
Furthermore, we have shown that in this regime, the number entropy $S_n$ gives a good estimate for the central charge.
The number entropy can be extracted from snapshots generated routinely by existing quantum gas microscopes.
Other experimentally accessible observables like on-site correlations, the momentum distribution and chiral currents confirm the transition from the trivial to the Laughlin phase.
Finally, we have identified the topological nature of the strong-coupling phase by extracting the many-body Chern number.

The system studied here, consisting of tunably coupled chains, provides a promising route towards the adiabatic preparation, detection and characterization of interacting topological states of matter using existing experimental techniques.
Measuring the entanglement entropy using more than one basis has been proposed~\cite{Bergh2021,Bergh2021a} and might give further insight into similar systems.

\begin{acknowledgments}
\ 

The authors would like to thank Monika Aidelsburger, Maximilian Buser, Markus Greiner, Mohammad Hafezi, Matja\v{z} Kebri\v{c}, Joyce Kwan, Julian L{\'{e}}onard, Sebastian Paeckel, Henning Schl\"omer for fruitful discussions. 
We acknowledge funding by the Deutsche Forschungsgemeinschaft (DFG, German Research Foundation) under Germany's Excellence Strategy -- EXC-2111 -- 390814868, and via Research Unit FOR 2414 under project number 277974659. AB acknowledges funding by the NSF through a grant for the Institute for Theoretical Atomic, Molecular, and Optical Physics at Harvard University and the Smithsonian Astrophysical Observatory.
\end{acknowledgments}

\bibliographystyle{apsrev4-2}
%

\newpage
\widetext
\newpage
\begin{center}
\textbf{\Large{Supplemental Material}}
\end{center}
\setcounter{equation}{0}
\setcounter{figure}{0}
\setcounter{table}{0}
\setcounter{page}{1}
\makeatletter
\renewcommand{\theequation}{S\arabic{equation}}
\renewcommand{\thefigure}{S\arabic{figure}}
\renewcommand{\bibnumfmt}[1]{[S#1]}

\section{Technicalities regarding DMRG simulations}
The numerical results in this work are obtained using the single-site variant of the density-matrix renormalization-group (DMRG) method~\cite{White1992,Schollwoeck2005} with subspace expansion~\cite{Hubig2015}.
We exploit $\mathrm{U}(1)$-symmetry associated with the particle number conservation to determine the variational ground state of the Hofstadter-Bose-Hubbard (HBH) model for a specified number of particles.
Convergence of the simulations is ensured by comparing the ground state energy $\left\langle\hat{\cal{H}}\right\rangle$ and the corresponding variance $\left\langle \hat{\cal{H}}^2 \right\rangle - \left\langle \hat{\cal{H}} \right\rangle^2$ for different bond dimensions up to $\chi = 6000$ for open boundary conditions and $\chi = 3000$ for periodic boundary conditions in one direction.

\begin{figure}[h]
	\centering
	\includegraphics[width=0.5\linewidth]{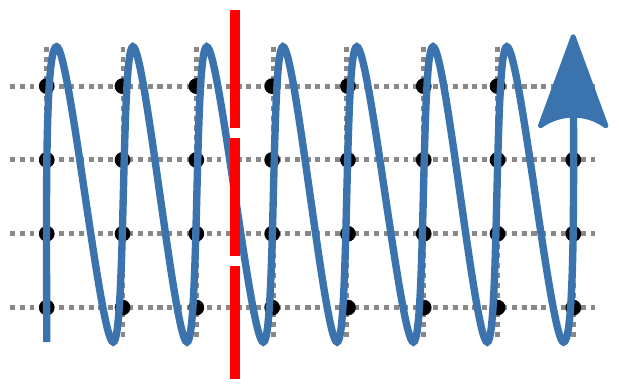}
	\caption{Given the lattice sites of the two-dimensional square lattice (black dots) the MPS chain (blue line) is chosen such that neighboring sites in $y$-direction are also neighboring sites in the MPS chain.
		Thus, a cut perpendicular to the chains in $x$-direction (red dashed line) can be realized by cutting a single bond in the MPS chain, giving immediate access to the bipartite entanglement entropy as function of the cut's position.}
	\label{Fig:Supp:MPS-Chain}
\end{figure}

The choice of the matrix product state (MPS) chain depicted in Fig.~\ref{Fig:Supp:MPS-Chain} allows for a simple evaluation of the bipartite entanglement entropy by cutting a single MPS bond.

When evaluating the many-body Chern number, exponential operators have to be applied to the ground state wave function~\cite{Dehghani2021}.
During this application the resulting MPS are truncated at bond dimension $\chi = 150$.
The necessary SWAP operation is implemented on the level of an MPS overlap.

\section{Extracting the central charge}
\begin{figure}
	\centering
	\includegraphics[width=0.49\linewidth]{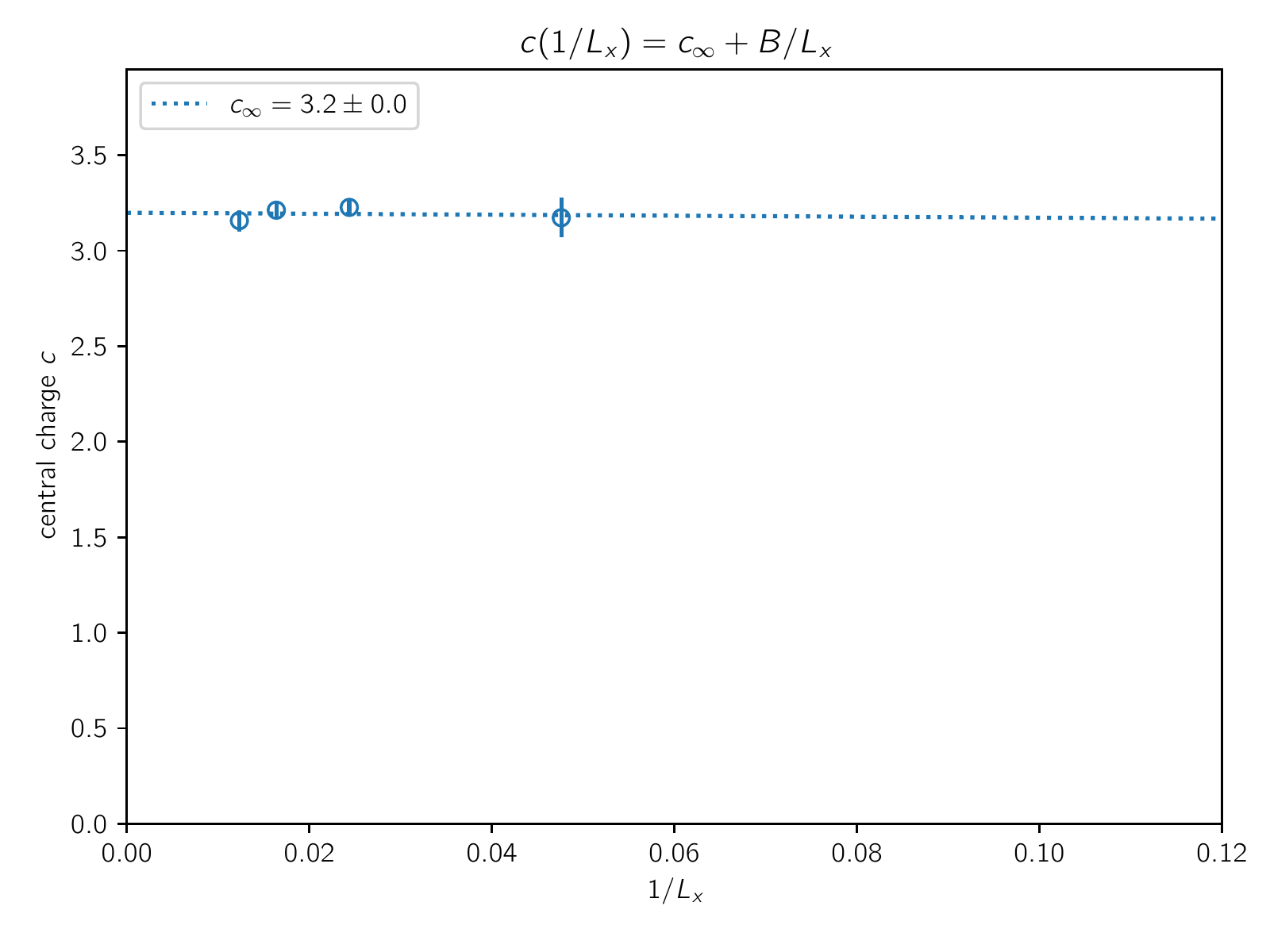}\hfill
	\includegraphics[width=0.49\linewidth]{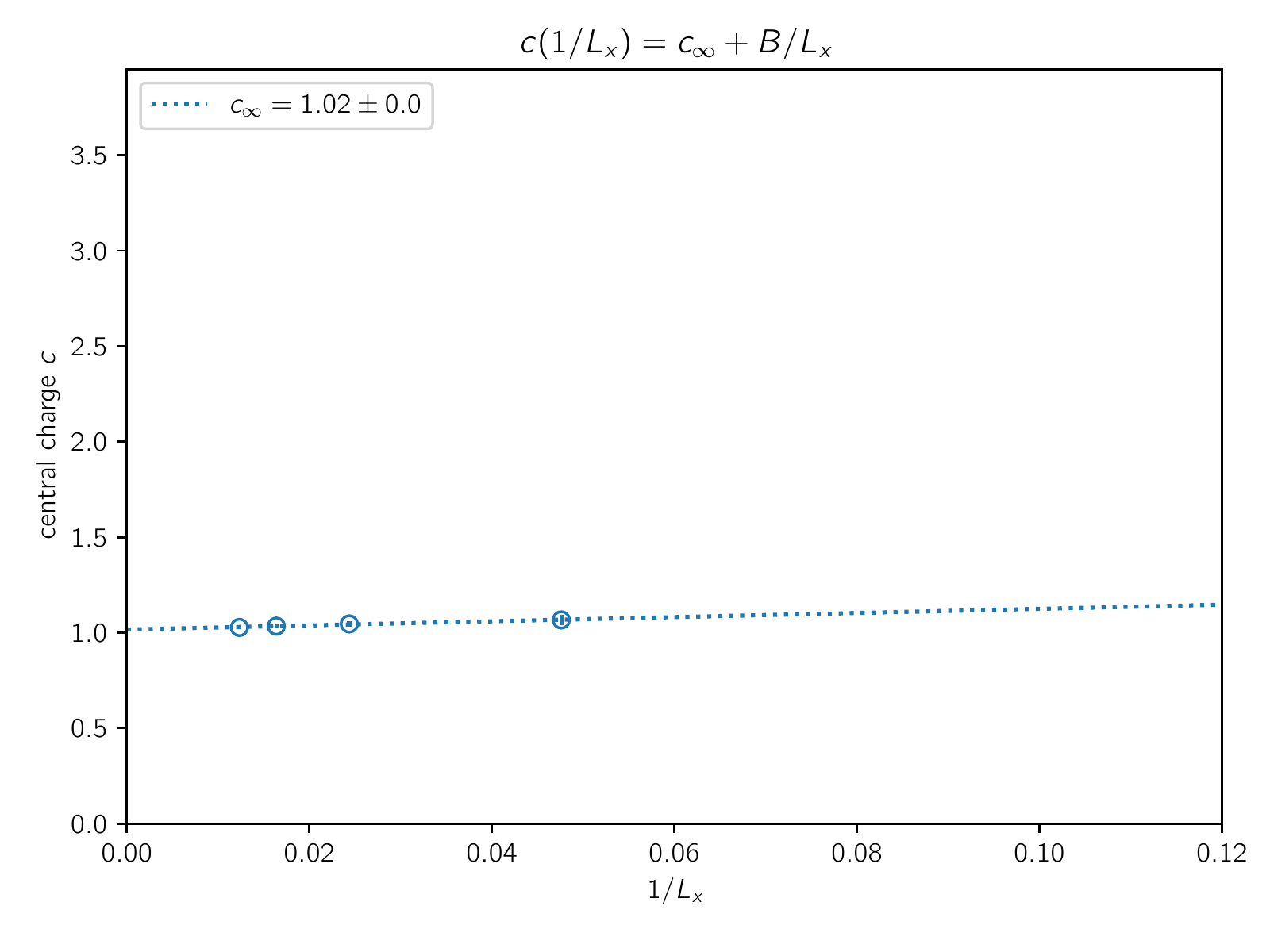}
	\caption{Extrapolation of the central charge from finite $L_x$ to $L_x\to\infty$ for $t_y/t_x = 0.2$ (left) and $t_y/t_x = 1.0$.
	The error bars of the single data points indicate the uncertainty of the fit with the CFT formula.}
	\label{Fig:Supp:ExtrapolationLx}
\end{figure}

As a function of the cuts position $\ell$, the bipartite entanglement entropy $S(\ell)$ shows significant Friedel-like oscillations in some cases.
To account for these oscillations, we normalize the entanglement entropy in a given bond $\ell$ by the local densities at the sites $x=\ell\pm 1/2$.
Defining the density
\begin{equation}
	n(x) = \frac{1}{L_y} \sum_{y=1}^{L_y} \left\langle\hat{n}_{x,y}\right\rangle,
\end{equation}
and the average density $\bar{n} = \frac{1}{L_x}\sum_{x=1}^{L_x}n(x)$, we therefore define the rescaled entanglement entropy
\begin{equation}
	\tilde{S}(\ell) = \frac{2S(\ell)}{\left(n(\ell-\frac{1}{2}) + n(\ell+\frac{1}{2})\right)} \bar{n}.
\end{equation}
In addition, to reduce the effect of the truncation at a finite bond dimensions $\chi$, we extrapolate the rescaled entanglement entropies $\tilde{S}(\ell; \chi)$to infinite bond dimensions as
\begin{equation}
	\tilde{S}(\ell; \chi) = \tilde{S}(\ell; \chi\to\infty) + A \log\left(1+\nicefrac{1}{\chi}\right).
\end{equation}
Next, the extrapolated entanglement entropies $\tilde{S}(\ell; \chi\to\infty)$ are fitted using the CFT prediction~\cite{Calabrese2004}
\begin{equation}
	S_{\rm CFT}(\ell) = \frac{c}{6} \log\left(\frac{2 L_x}{\pi} \sin\left(\frac{\pi \ell}{L_x}\right)\right) + g,
\end{equation}
where $c$ is the central charge and $g$ is some non-universal constant.
Finally, we extrapolate the finite-size values for the central to infinite systems by a first-order extrapolation in $\nicefrac{1}{L_x}$,
\begin{equation}
	c(L_x) = c(L_x\to \infty) + \nicefrac{B}{L_x}.
\end{equation}
Examples for this extrapolation can be found in Fig.~\ref{Fig:Supp:ExtrapolationLx}.

\begin{figure}
	\centering
	\includegraphics[width=0.49\linewidth]{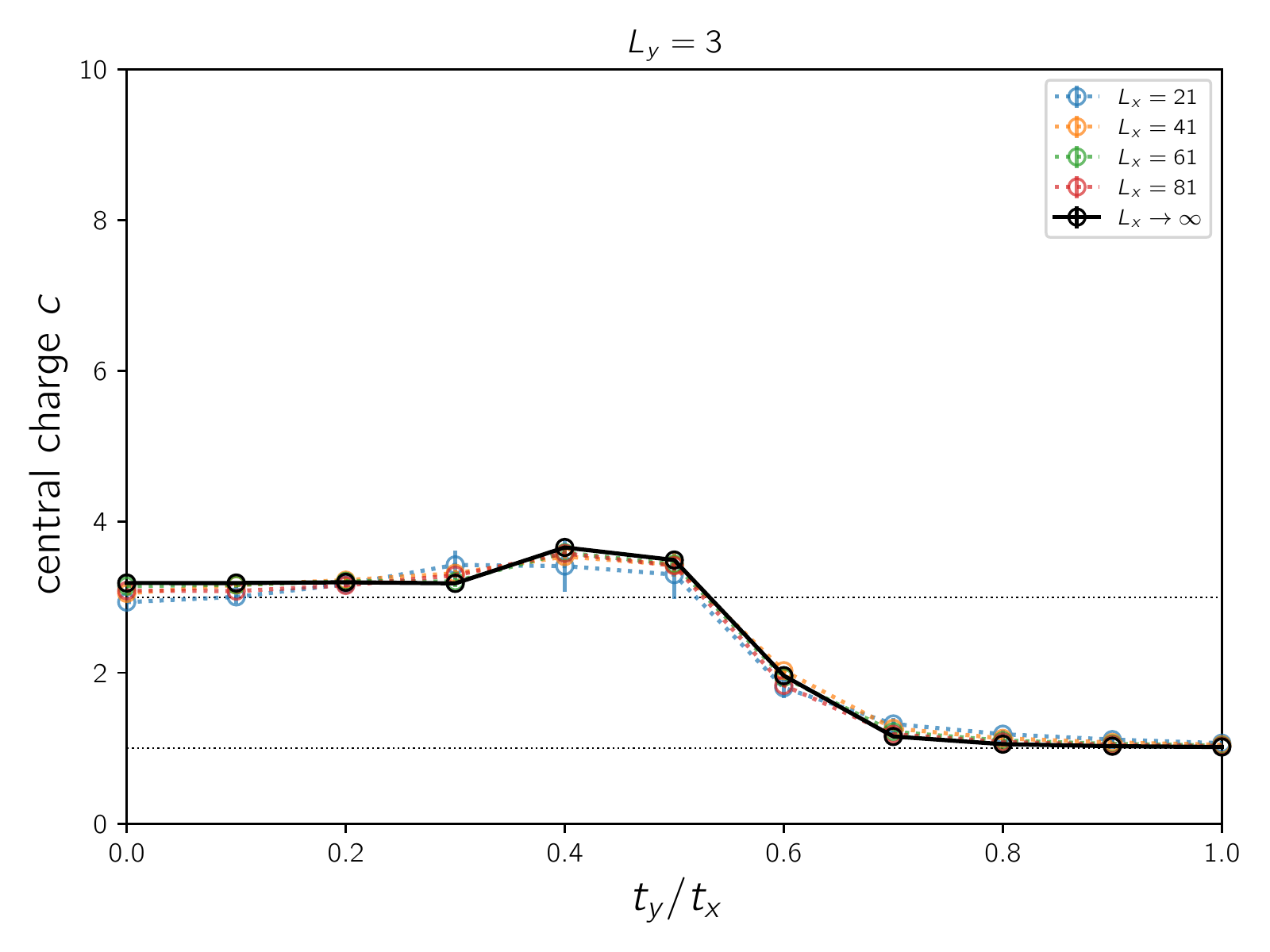}
	\includegraphics[width=0.49\linewidth]{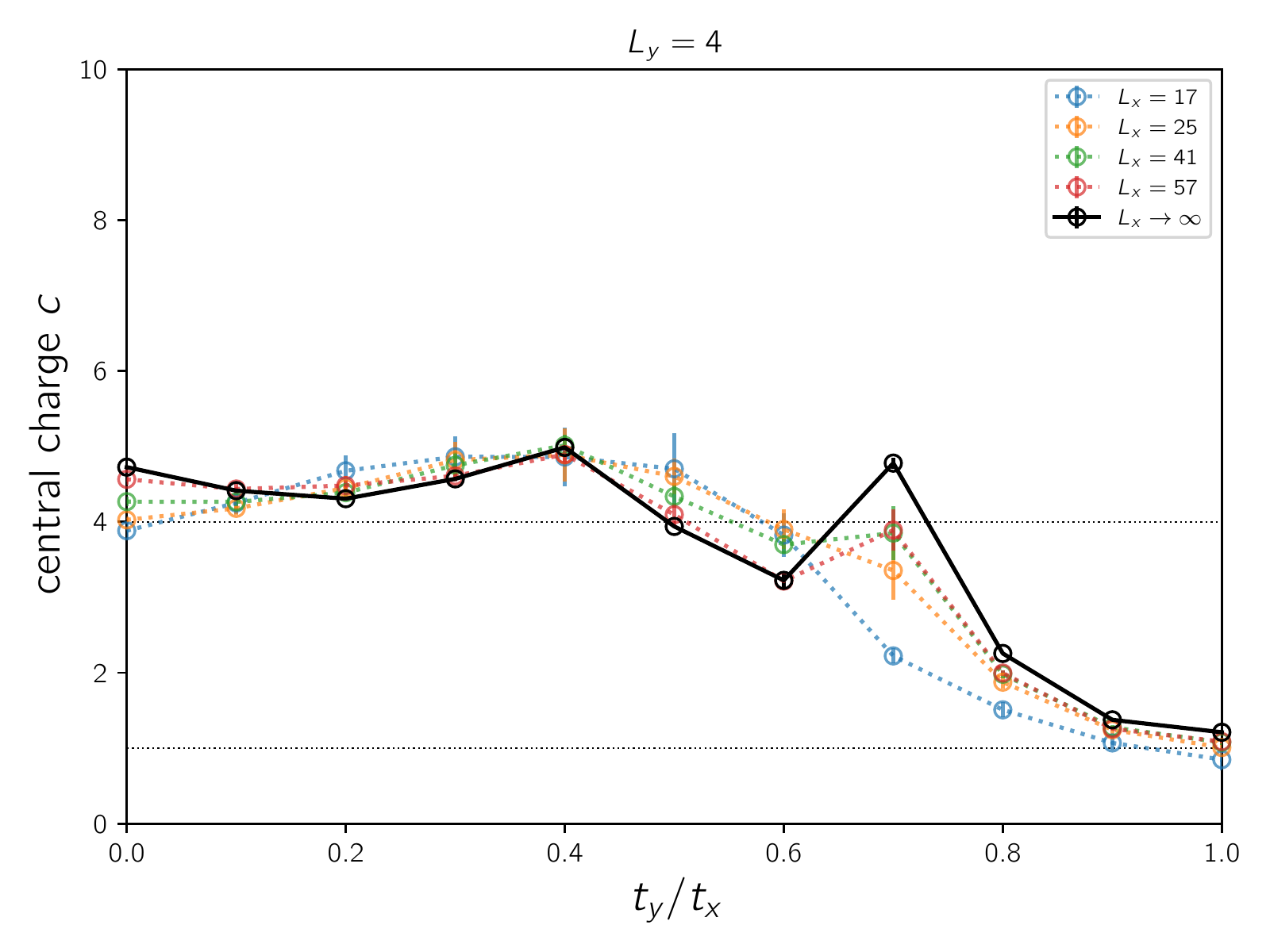}
	\includegraphics[width=0.49\linewidth]{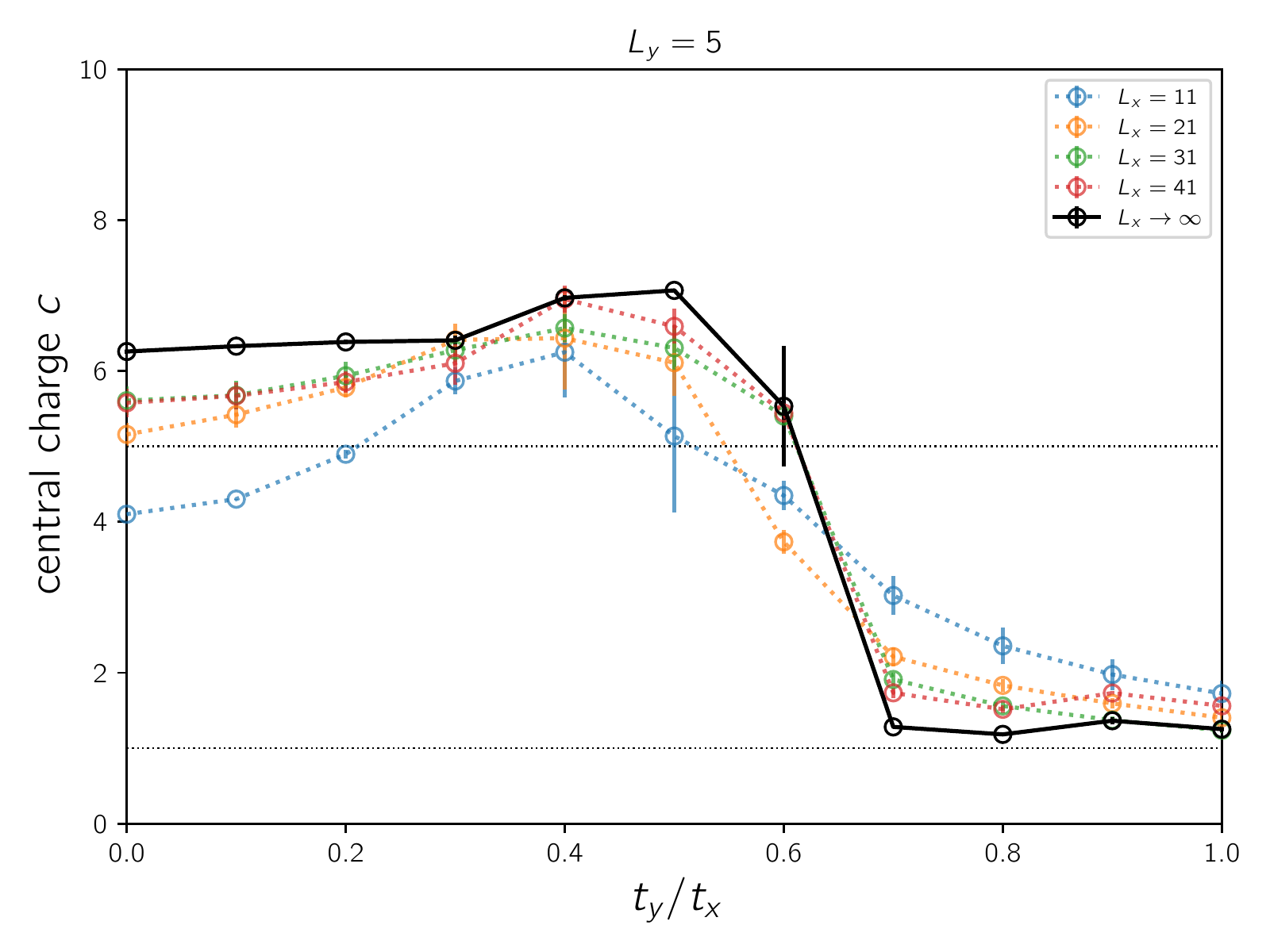}
	\includegraphics[width=0.49\linewidth]{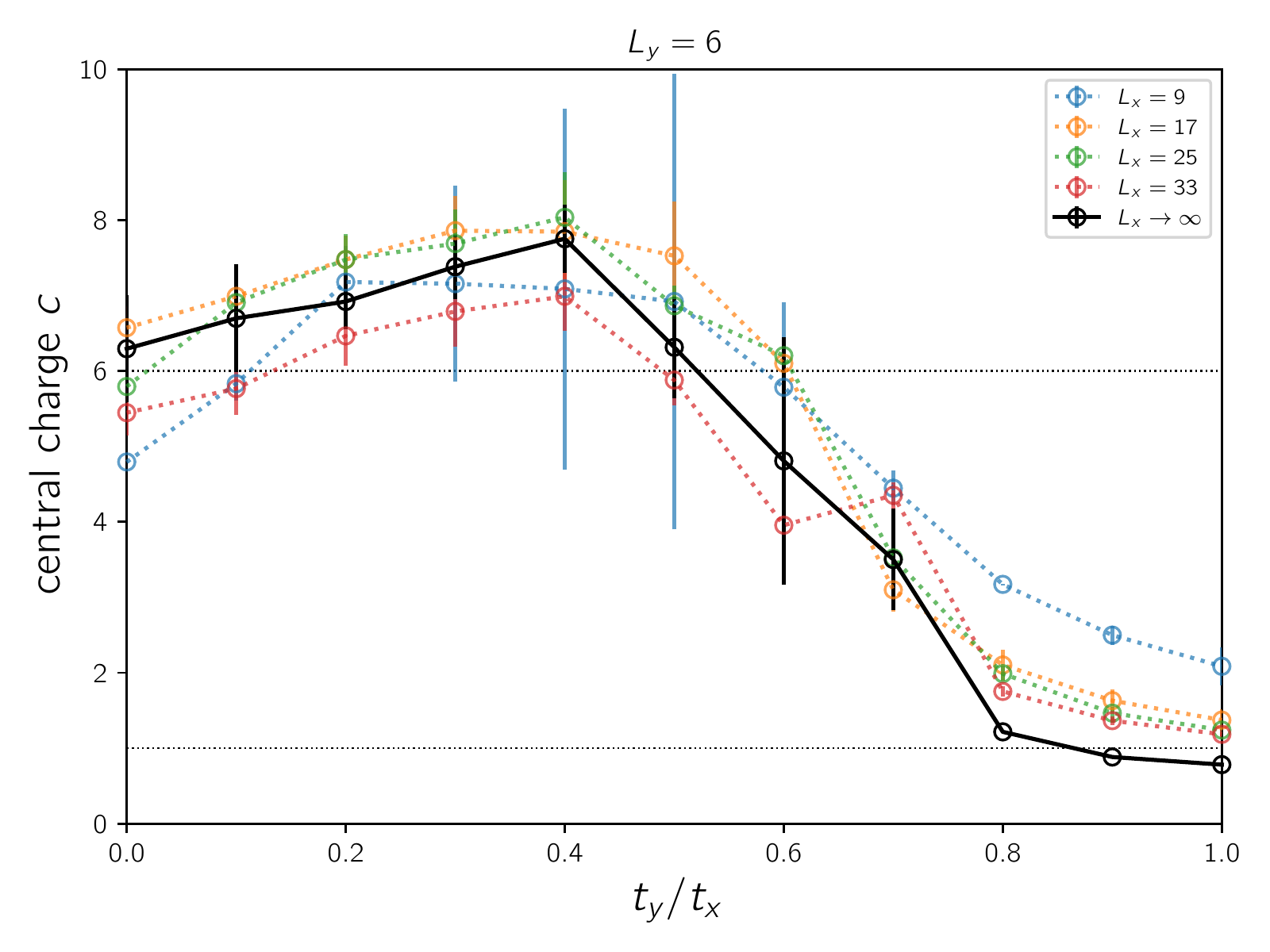}
	\caption{Central charge as obtained from the rescaled and $\chi$-extrapolated bipartite entanglement entropy for different system sizes ($L_y=3,..., 6$) and the extrapolated value in the limit $L_x \to \infty$.
	The dotted lines indicate $c=1, L_y$ respectively.}
	\label{Fig:Supp:CentralCharge}
\end{figure}

We show the results for the central charge before the extrapolation to infinite systems in Fig.~\ref{Fig:Supp:CentralCharge}.
Especially in the Laughlin regime at large $t_y/t_x$ we find excellent convergence of the central charge to the predicted value $c_{\rm LN}=1$ in all systems.

\section{Chiral edge current}
\begin{figure}
	\centering
	\includegraphics[width=0.3333\linewidth]{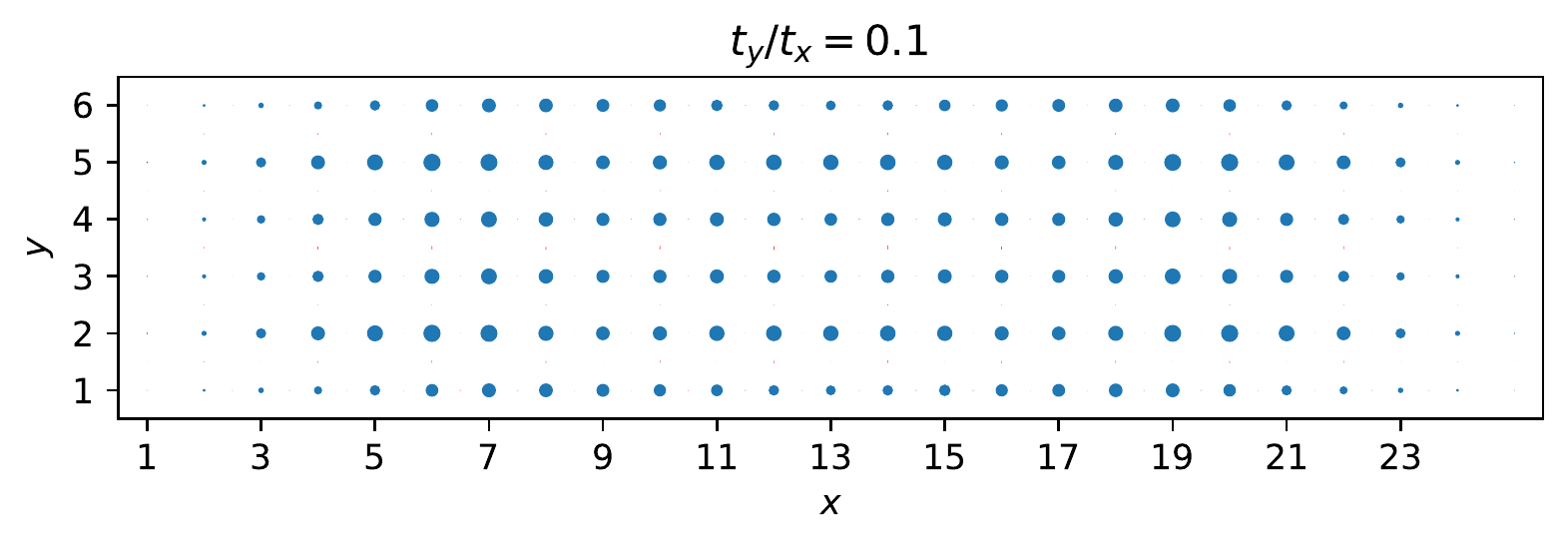}\hfill
	\includegraphics[width=0.3333\linewidth]{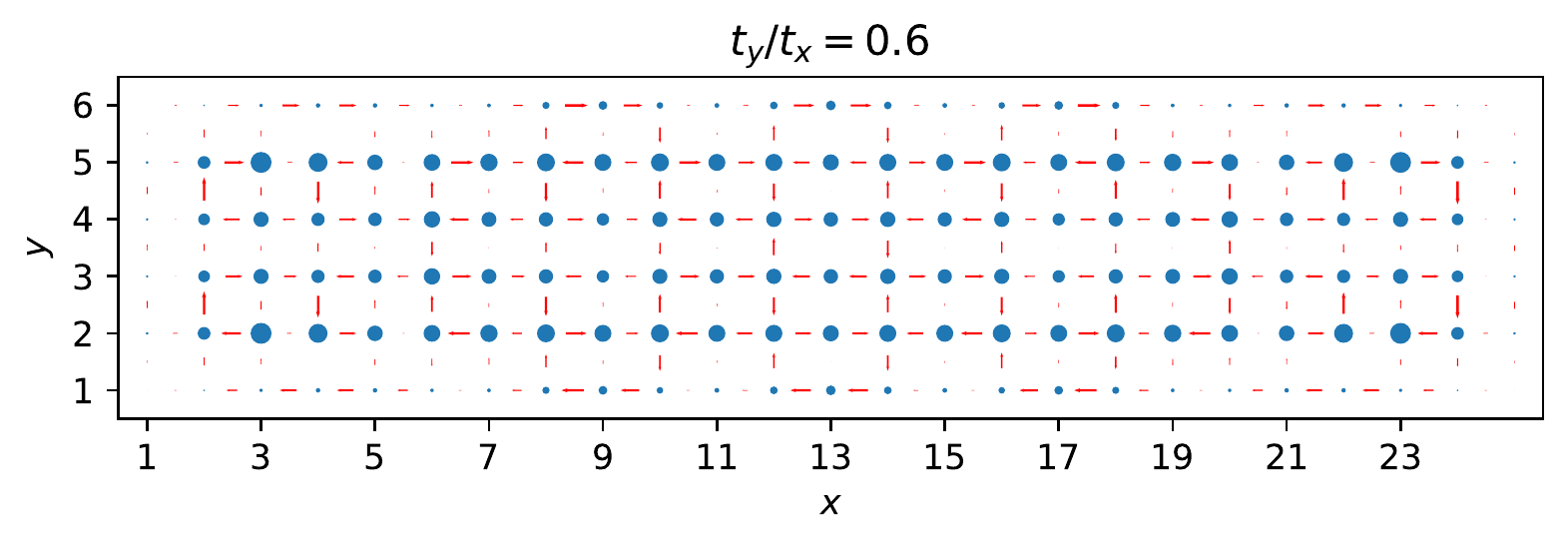}\hfill
	\includegraphics[width=0.3333\linewidth]{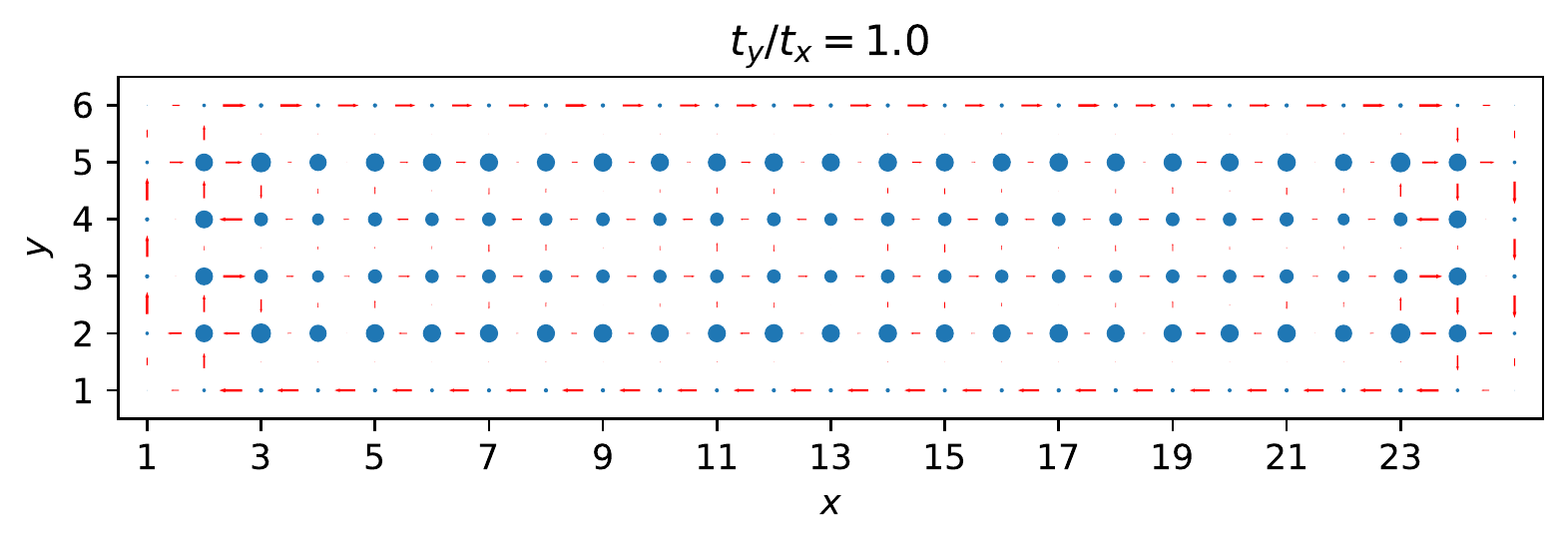}
	\caption{Local currents for a system of size $L_x = 25, L_y=6$ and $N=15$ particles subject to $N_{\phi}=30$ flux quanta.
		The length and width of the red arrows indicate the strength and direction of the current along a bond and the radii of the blue dots indicate the density at a given site.
		The hopping strength in the $y$-direction is increased from left to right, $t_y/t_x = 0.1, 0.6, 1.0$.
		Note the pronounced edge current in the isotropic limit.}
	\label{Fig:CurrentPatterns}
\end{figure}

We define the currents in $x$- and $y$-direction as
\begin{align}
	j^x_{x,y} &= \left\langle\hat{j}^x_{x,y}\right\rangle = it_x \left\langle\hat{a}_{x+1, y}^{\dagger} \hat{a}_{x, y}^{\vphantom\dagger}\right\rangle + \text{c.c.},\\
	j^y_{x,y} &= \left\langle\hat{j}^y_{x,y}\right\rangle = it_y \mathrm{e}^{2\pi i\alpha x} \left\langle\hat{a}_{x, y+1}^{\dagger} \hat{a}_{x, y}^{\vphantom\dagger}\right\rangle + \text{c.c.}.
\end{align}
In the regime of weakly coupled wires we expect no pronounced edge current, while in the Laughlin state the edge current should be far more prominent than the current fluctuations in the bulk.
Typical current patterns are depicted in Fig.~\ref{Fig:CurrentPatterns} for the weakly coupled, the transient, and the Laughlin regime.
It is clearly visible that above some critical value of $t_y/t_x$ a chiral edge current forms.

\section{Calculating the many-body Chern number}
For the calculation of the many-body Chern number we perform DMRG calculations on cylinders of coupled chains.
Following~\cite{Dehghani2021}, we choose three cylindrical regions $R_{1,2,3}$ of length $\ell_x=\lfloor (L_x-3)/ 3\rfloor$ on the cylinder, where $\lfloor x\rfloor$ denotes the integer part of $x$, and define
\begin{equation}
	\mathcal{T}(\phi, s) = \left\langle\hat{W}_{R_1}^{\dagger}(\phi) \hat{S}_{1,3}\hat{W}_{R_1}(\phi)\hat{V}_{R_1\cup R_2}^s\right\rangle,
\end{equation}
where $\hat{S}_{1,3}$ swaps the regions $R_1$ and $R_3$ and
\begin{equation}
	\hat{W}_{R}(\phi) = \prod_{(x,y)\in R} \mathrm{e}^{i\hat{n}_{x,y}\phi}, \text{ and }
	\hat{V}_{R} = \prod_{(x,y)\in R}\mathrm{e}^{i\frac{2\pi y}{L_y}\hat{n}_{x,y}}
\end{equation}
are defect operators.
In our case, $s$ is the ground state degeneracy on a torus, which for the $\nicefrac{1}{2}$-Laughlin state is known to be $s=2$.
Varying $\phi$ from $0$ to $2\pi$, we obtain the many-body Chern number as the winding number
\begin{equation}
	C = \frac{1}{2\pi} \oint \mathrm{d}\phi\ \frac{\mathrm{d}}{\mathrm{d}\phi} \arg\mathcal{T}(\phi, s),
\end{equation}
which is directly related to the Hall conductivity \mbox{$\sigma_{\rm H} = \frac{C}{s} \frac{e^2}{h}$}.
\begin{figure}
	\centering
	\includegraphics[width=0.5\linewidth]{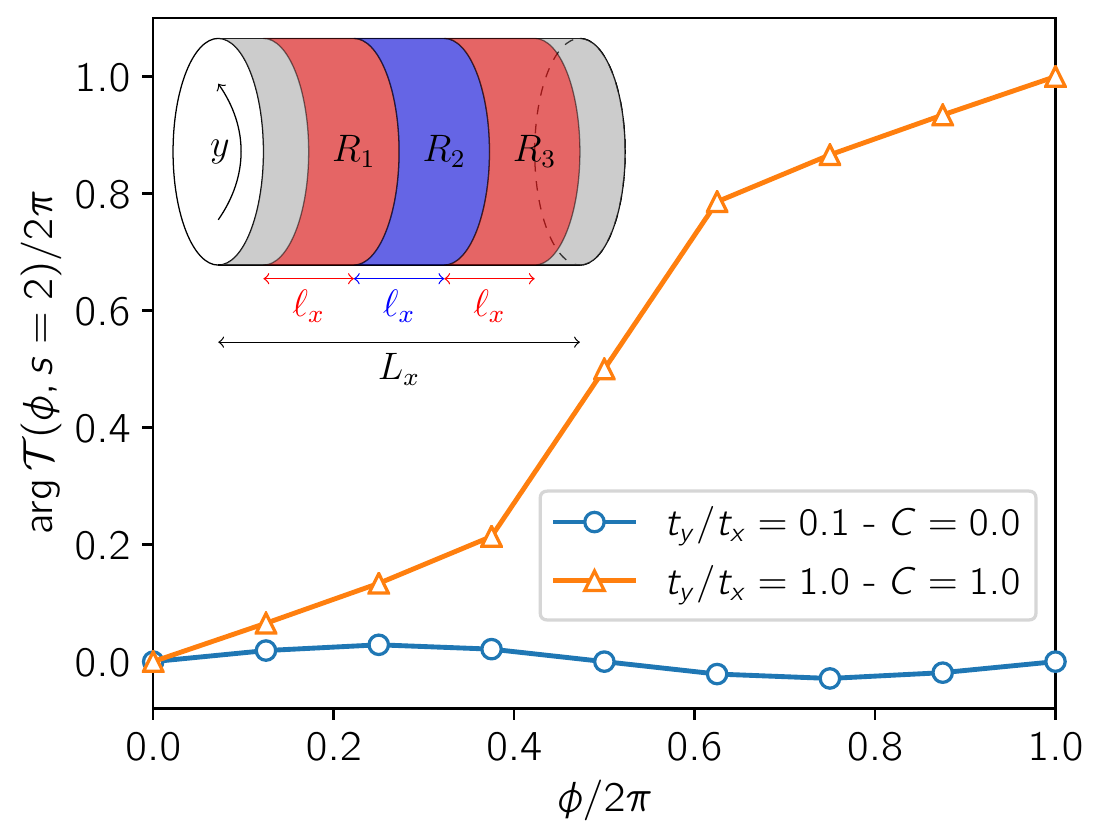}
	\caption{Origin of different Chern numbers from the winding of $\mathcal{T}(\phi, s=2)$ in the 	complex plane.
		The sketch illustrates the regions used for the evaluation of the Chern number.}
	\label{Fig:Supp:Chern}
\end{figure}

This result also allows us to confirm our earlier assumption for the value of $s$ by comparing $\sigma_H$ obtained by this method with the Hall conductivity obtained from a charge pump procedure.
To this end, we introduce a magnetic flux $\phi$ through the cylinder by modifying the Hamiltonian of the model to be
\begin{equation}
	\hat{\mathcal{H}}\left(\phi\right) = - t_x \sum_{x=1}^{L_x-1} \sum_{y=1}^{L_y} \left(\hat{a}^{\dagger}_{x+1, y}\hat{a}^{\vphantom{\dagger}}_{x,y} + \mathrm{H.c.}\right) - t_y \sum_{x=1}^{L_x} \sum_{y=1}^{L_y-1} \left(\mathrm{e}^{2\pi i \alpha x + i \phi x/L_y}\hat{a}^{\dagger}_{x,y+1}\hat{a}^{\vphantom{\dagger}}_{x,y} + \mathrm{H.c.}\right) +\frac{U}{2}\sum_{x,y} \hat{n}_{x,y}\left(\hat{n}_{x,y}-1\right).
\end{equation}
We perform DMRG ground state searches for slowly increasing value of $\phi$ with maximum bond dimension $\chi = 200$ to simulate an adiabatic flux insertion protocol.
We cut the cylinder into two halves and track the particle number $n_L(\phi)$ in the left half as function of the inserted flux $\phi$.
Taking into account the unit charge of the bosons, we can define the pumped charge as
\begin{equation}
	\delta Q(\phi) = n_L(\phi) - n_L(0).
\end{equation}

\begin{figure}
	\centering
	\includegraphics[width=0.5\linewidth]{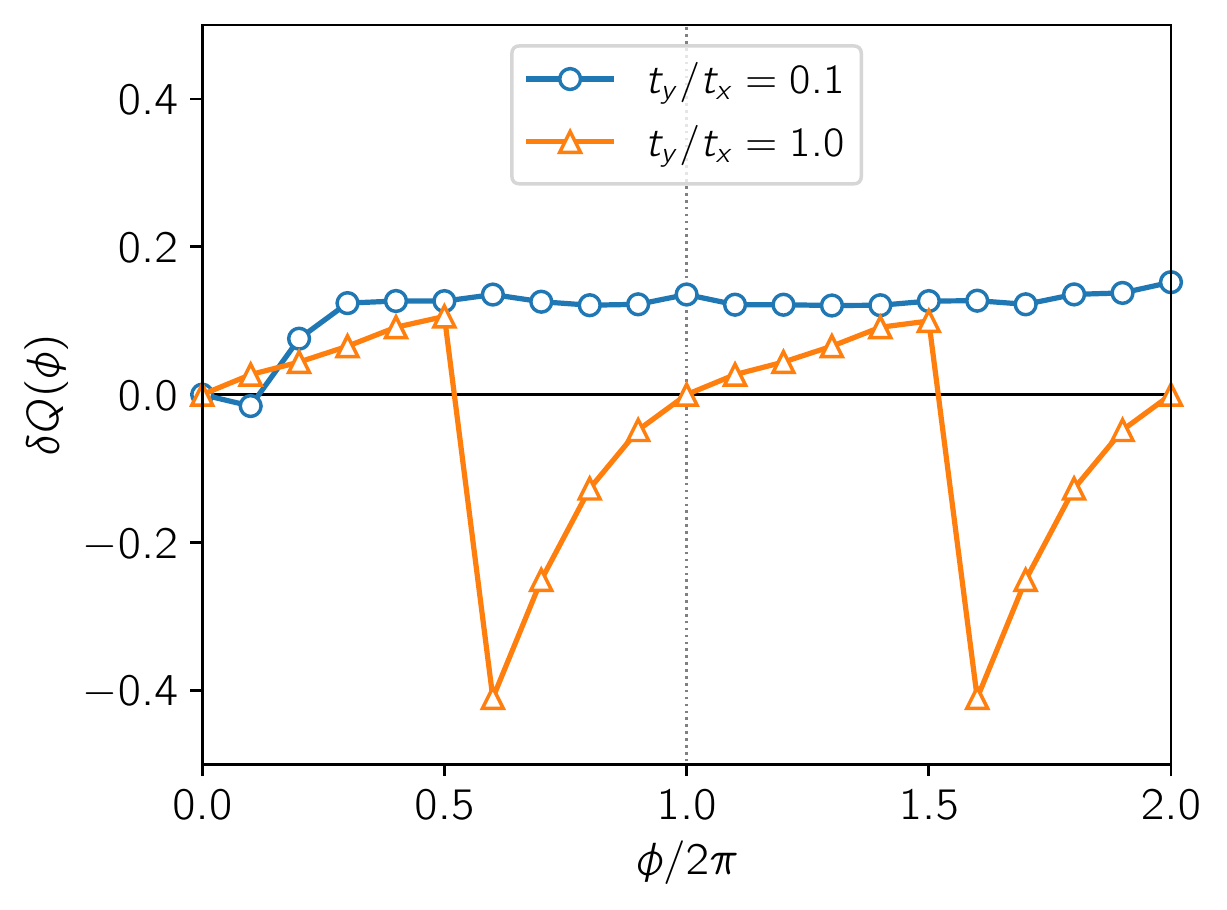}
	\caption{Pumped charge as function of inserted flux for two different limits of the interchain coupling.
	%
	%
		In the weakly coupled limit, $t_y/t_x=0.1$, only a small, non-significant amount of charge is pumped.
		We attribute the remaining pumped charge to convergence difficulties in this regime.
		In the isotropic limit, $t_y/t_x=1.0$, an effective unit charge is pumped after the insertion of $4\pi$ flux.
		The jumps are due to the charge density wave forming on thin cylinders jumping to a competing configuration with occupied orbitals shifted by one.}
	\label{Fig:Supp:PumpedCharge}
\end{figure}

The results of this procedure are visualized in Fig.~\ref{Fig:Supp:PumpedCharge} for two values of the interchain hopping.
In particular, we find that in the weakly coupled limit, $t_y/t_x = 0.1$, no significant charge is pumped and therefore no Hall response is observed.
We attribute the small residual pumped charge to the difficult convergence in this challenging regime.
In contrast, in the isotropic limit, $t_y/t_x = 1.0$, we find a pumped unit charge after inserting flux $\phi=2\times 2\pi$.
This is consistent with a Hall conductivity of $\sigma_H^{\rm pump} = \frac{1}{2} \frac{e^2}{h}$ after restoring units.
This clearly shows that $s=2$ is indeed the right ground state degeneracy in the topological regime and a similar approach can be used to obtain the correct factor also for more involved cases.

\section{$L_y = 2$ - Bosonic flux ladders}
\begin{figure}
	\centering
	\includegraphics[width=0.49\linewidth]{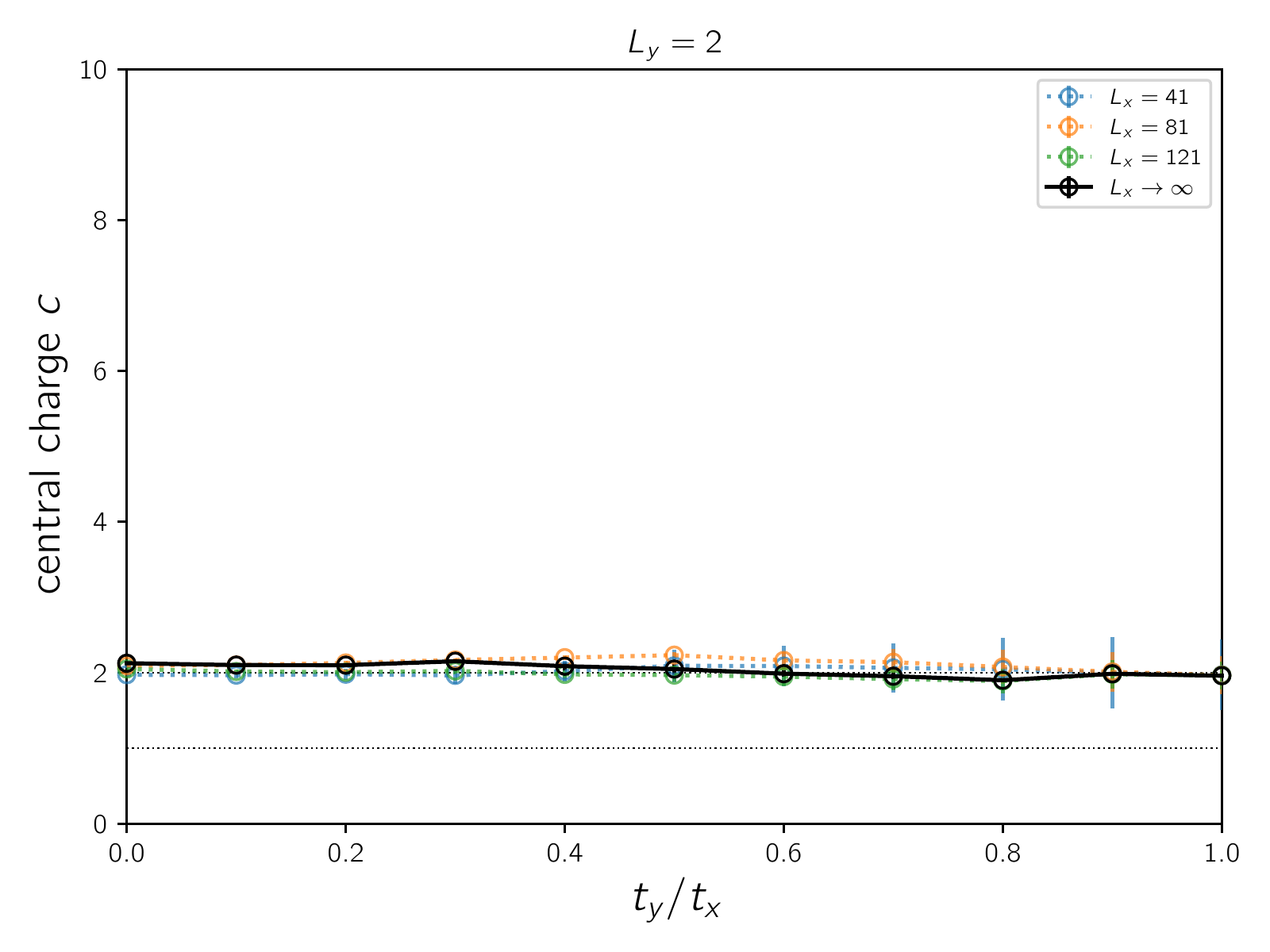}
	\caption{Extracted central charge for ladder systems with $L_y=2$.
		Note the qualitative difference from the extended systems in the main text, in particular the absence of a transition toward $c_{\rm LN}=1$ (indicated by the dotted line).}
	\label{Fig:Supp:Ladder:CentralCharge}
\end{figure}

\begin{figure}
	\centering
	\includegraphics[width=0.49\linewidth]{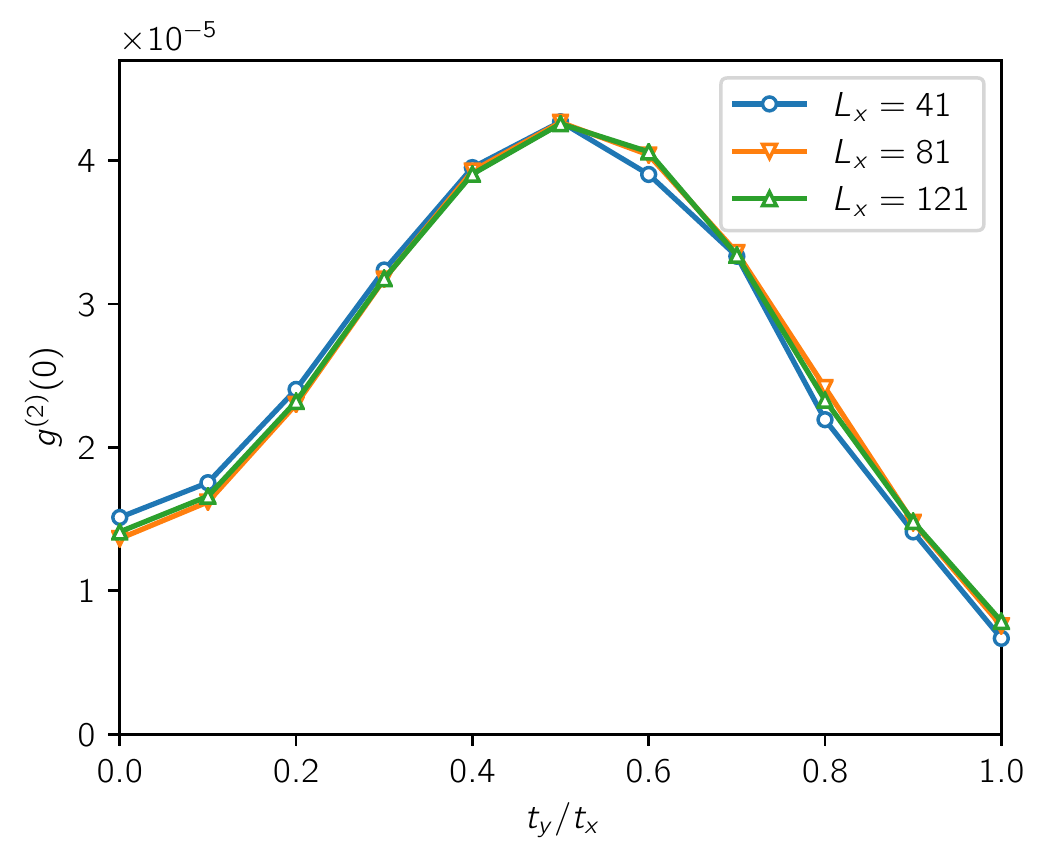}
	\includegraphics[width=0.49\linewidth]{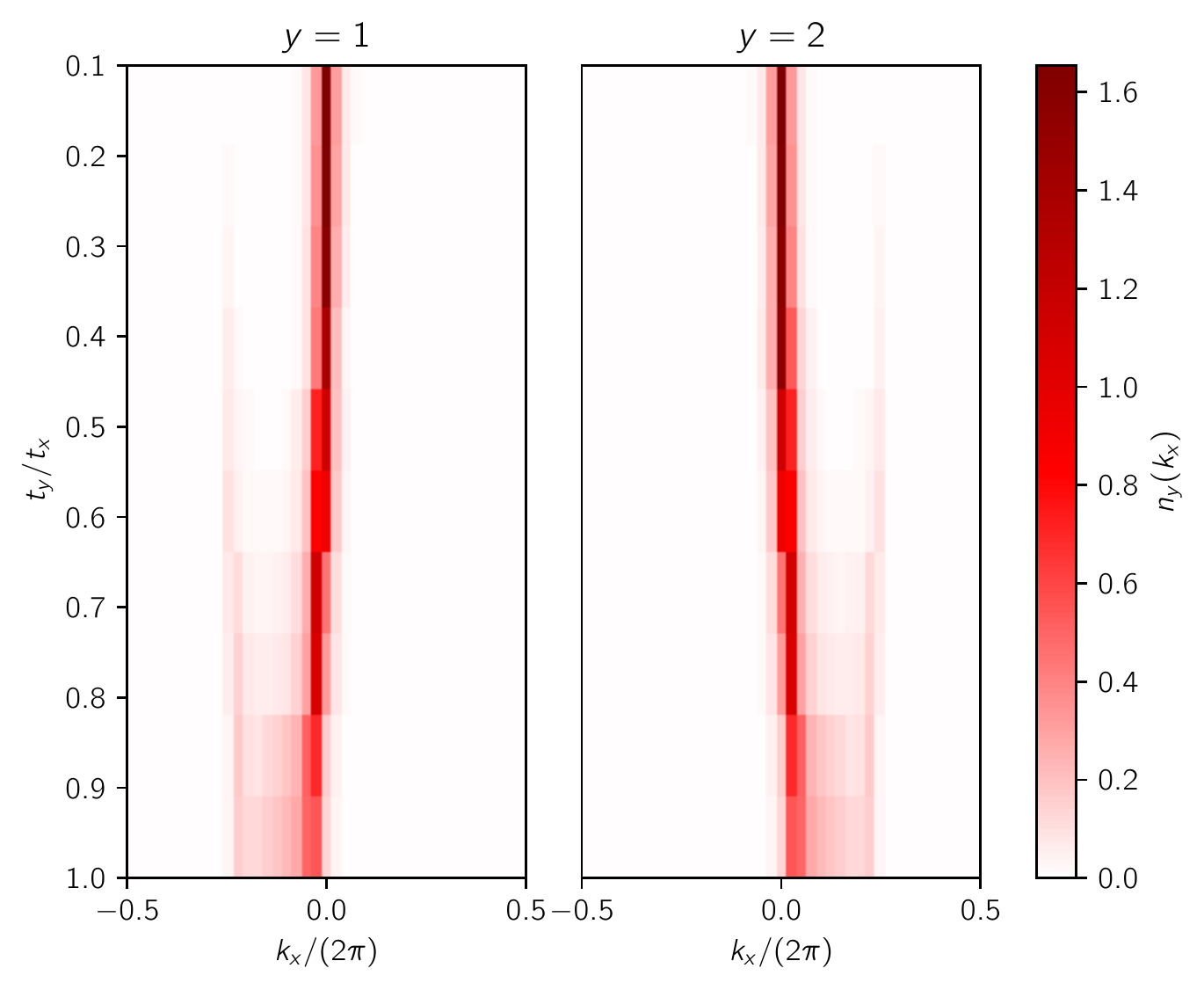}
	\caption{On-site correlations (left) and momentum distribution in the two legs (right) for a ladder system.
		There are some signatures reminiscent of the features in larger systems.}
	\label{Fig:Supp:Ladder:OtherObservables}
\end{figure}

As a limiting case we simulated two-leg ladders ($L_y=2$) subject to a perpendicular magnetic field.
Similar models were studied with various modifications in earlier studies~\cite{Huegel2014,Petrescu2015,Piraud2015,DiDio2015,Strinati2017,Petrescu2017,Strinati2019,Buser2020,Buser2021}.
We find that the physics in this regime is qualitatively different from the extended systems with $L_y \geq 3$ studied in the main text.
The extrapolated central charge for the ladder system is depicted in Fig.~\ref{Fig:Supp:Ladder:CentralCharge}.
Note, that we do not find a transition to the predicted $c_{\rm LN}=1$ for the Laughlin state, but instead find $c=2$ independent of the inter-chain hopping.
Nevertheless, the on-site correlations and the momentum distribution show some signatures reminiscent of larger systems, see Fig.~\ref{Fig:Supp:Ladder:OtherObservables}.

\section{Additional data for $L_y=4,6$}
\begin{figure}
	\centering
	\includegraphics[width=\linewidth]{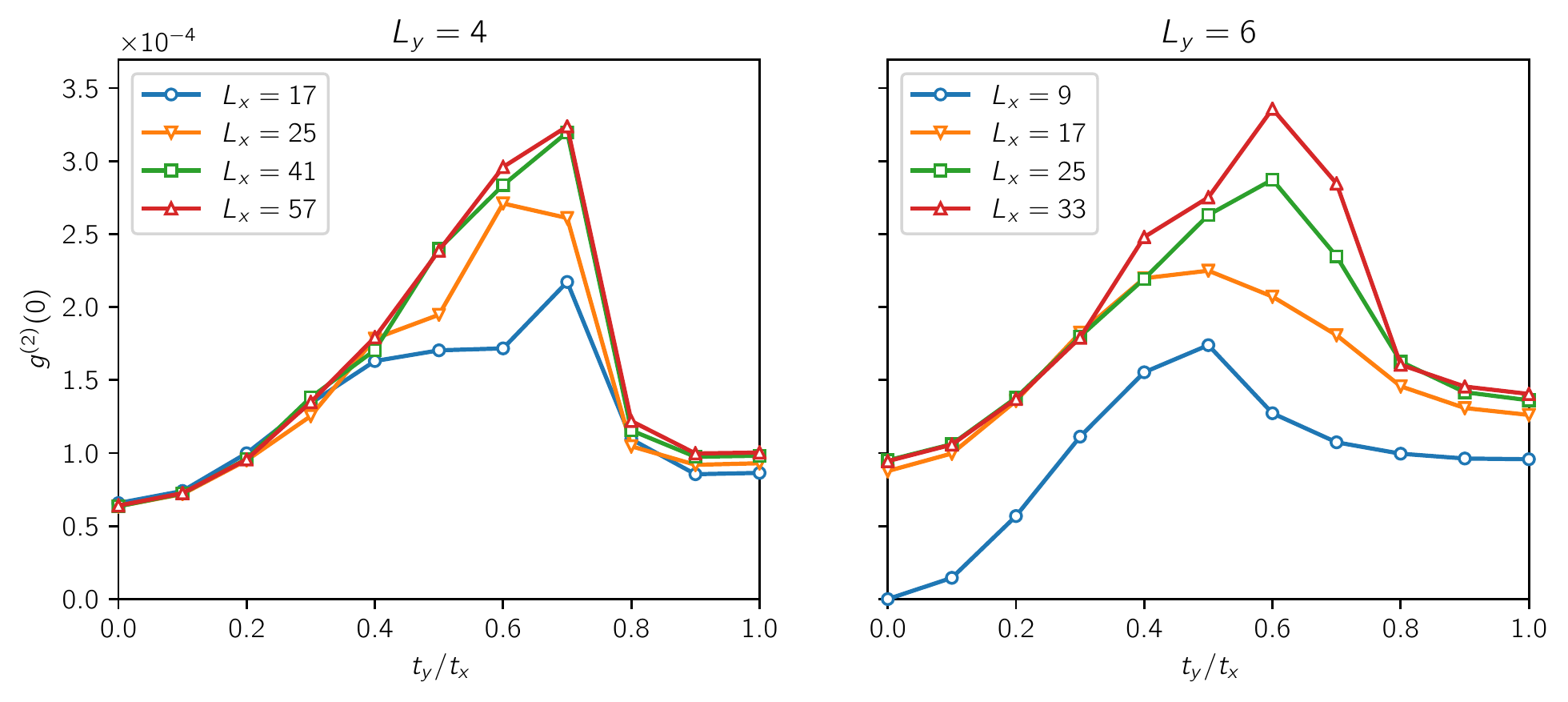}
	\caption{On-site correlations for $L_y=4$ and $L_y=6$ legs sharing the essential features discussed in the main text.}
	\label{Fig:Supp:HubbardLyEven}
\end{figure}

\begin{figure}
	\centering
	\includegraphics[width=0.49\linewidth]{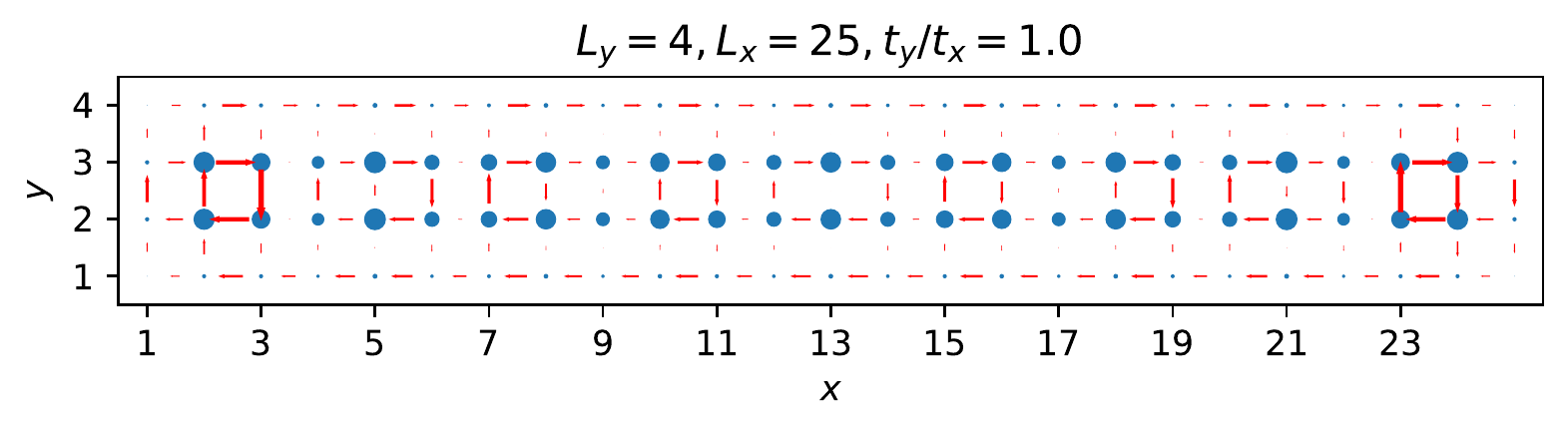}\hfill
	\includegraphics[width=0.49\linewidth]{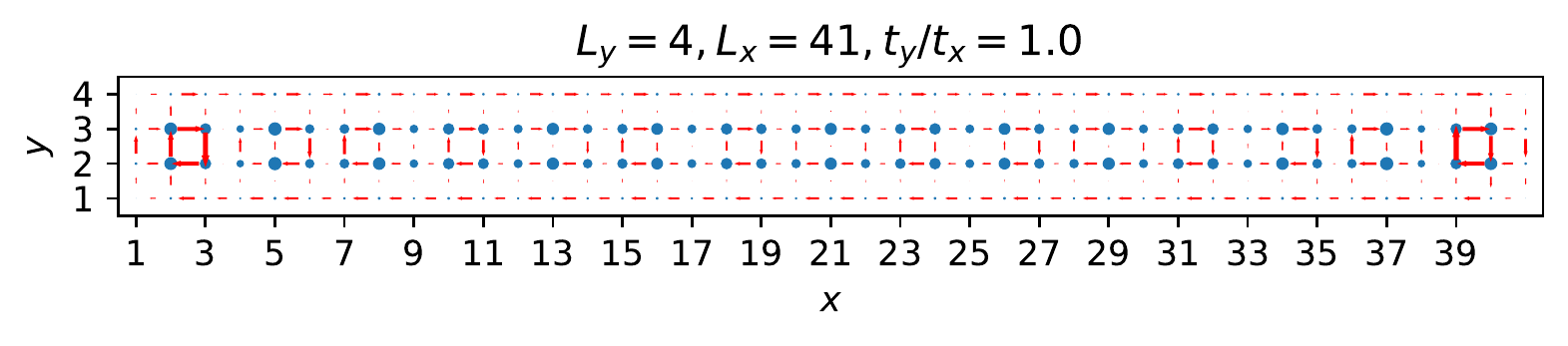}\\
	\includegraphics[width=0.49\linewidth]{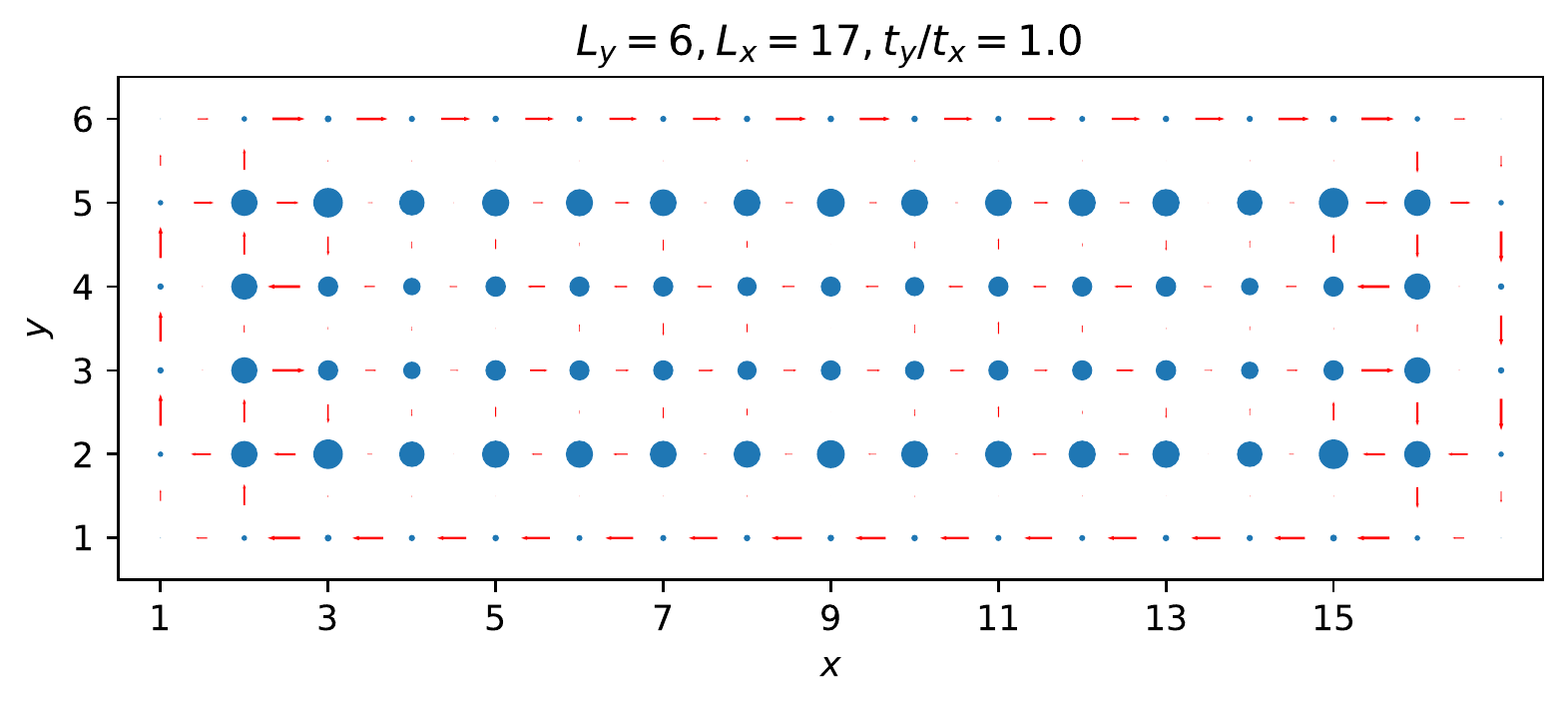}\hfill
	\includegraphics[width=0.49\linewidth]{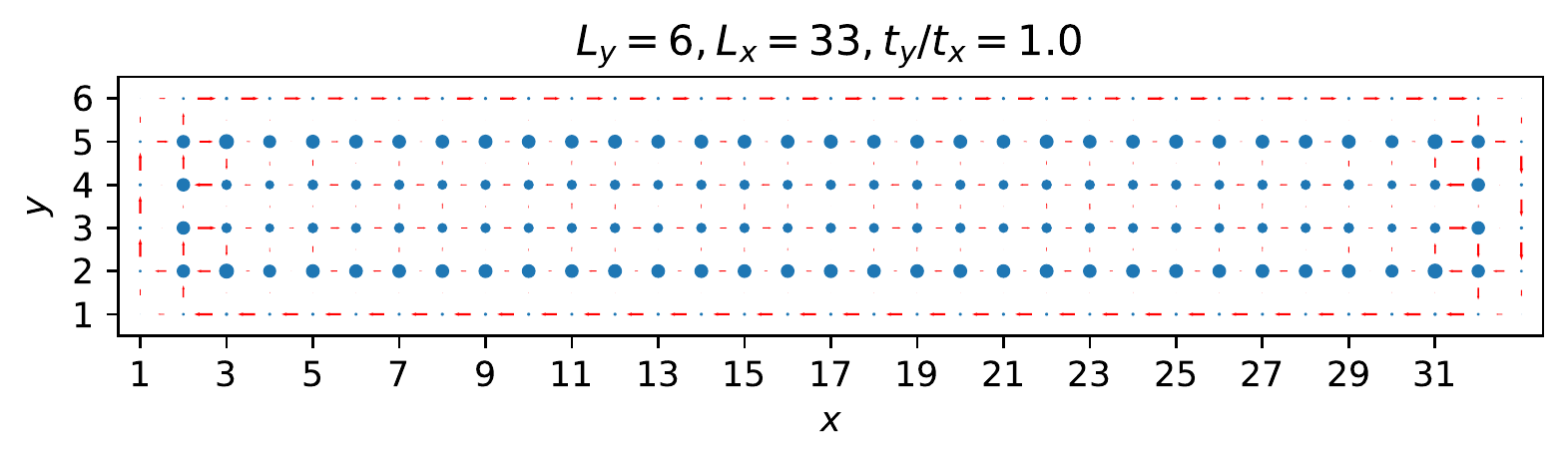}\\
	\caption{Vortices in the current pattern for even numbers of chains.
		For $L_y=6$ they are less pronounced, however the still differ from the current patterns observed for an odd number of legs (see Fig.~\ref{Fig:CurrentPatterns}).}
	\label{Fig:Supp:Vortices}
\end{figure}

For even numbers of chains, $L_y=4,6$, we find results which share the essential drop of the central charge close to the isotropic limit, see Fig.~\ref{Fig:Supp:CentralCharge}.
Also the on-site correlations share the essential features discussed for $L_y=3, 5$ in the main text, see Fig.~\ref{Fig:Supp:HubbardLyEven}.
However, in Fig.~\ref{Fig:Supp:Vortices} we find some non-trivial vortex pattern in the current for systems with an even number of chains which are less pronounced in the systems with $L_y=6$ chains.

\end{document}